\documentclass[11pt,preprintnumbers,superscriptaddress,nofootinbib,aps]{revtex4-1}
\usepackage{hyperref}
\usepackage{amssymb, amsmath}
\usepackage{color}

\newcommand{\lP}{\ensuremath{l_P}} 

\newcommand{\be}{\begin{equation}}
\newcommand{\ee}{\end{equation}}
\newcommand{\bea}{\begin{eqnarray}}
\newcommand{\eea}{\end{eqnarray}}
\begin{document}

\preprint{UTTG-05-16}

\title{Holographic entanglement chemistry}

\author{Elena Caceres}
\affiliation{Facultad de Ciencias, Universidad de Colima, Bernal Diaz del Castillo 340, Colima, Mexico}
\affiliation{Theory Group, Department of Physics, University of Texas, Austin, Texas 78712, USA}
\author{Phuc H. Nguyen}
\affiliation{Theory Group, Department of Physics, University of Texas, Austin, Texas 78712, USA}
\affiliation{Texas Cosmology Center, University of Texas, Austin, TX 78712, USA}
\author{Juan F. Pedraza}
\affiliation{Institute for Theoretical Physics, University of Amsterdam, 1090 GL Amsterdam, Netherlands}

\begin{abstract}
We use the Iyer-Wald formalism to derive an extended first law of entanglement that includes variations in the cosmological constant, Newton's constant and --in the case of higher-derivative theories--  all the additional couplings of the theory. In Einstein gravity, where the number of degrees of freedom $N^2$ of the dual field theory is  a function of $\Lambda$ and $G$, our approach allows us to  vary $N$ keeping  the field theory scale fixed or to vary the field theory scale keeping $N$ fixed. We also derive an extended first law of entanglement for Gauss-Bonnet and Lovelock gravity, and show that in these cases all the extra variations reorganize nicely in terms of the central charges of the theory. Finally, we comment on the implications for renormalization group flows and $c$-theorems in higher dimensions.
\end{abstract}

\maketitle

\section{Introduction}

In recent years, the notion of entanglement has played a crucial role in our understanding of quantum gravity and the emergence of spacetime.
Starting with Jacobson's seminal paper \cite{Jacobson:1995ab}, there have been several attempts to obtain gravitational dynamics from an underlying thermodynamical description,
with various degrees of success \cite{Padmanabhan:2003gd,Cai:2005ra,Eling:2006aw,Padmanabhan:2009vy,Verlinde:2010hp,Bianchi:2012ev,Lashkari:2013koa,Faulkner:2013ica,Jacobson:2015hqa}. This was in part motivated by the early work on black hole thermodynamics \cite{Bekenstein:1973ur,Bardeen:1973gs,Hawking:1974sw} and strongly supported
by the holographic principle, proposed by 't Hooft \cite{'tHooft:1993gx} and promoted by Susskind in \cite{Susskind:1994vu}. The discovery of the the AdS/CFT or gauge/gravity correspondence \cite{Maldacena:1997re} made it possible to frame some of these questions in more robust physical grounds, and has already proven to be a powerful arena to uncover deep connections between entanglement and gravity \cite{Swingle:2009bg,VanRaamsdonk:2009ar,VanRaamsdonk:2010pw,Czech:2012be,Swingle:2012wq}.

According to the AdS/CFT dictionary, black hole solutions in anti-de Sitter (AdS) are dual to strongly coupled large-$N$ gauge theories at finite temperature. Hence, in this context, black hole thermodynamics can be understood in terms of the fundamental degrees of freedom of a thermal quantum field theory and vice versa. For instance, the first law of thermodynamics maps to a bulk equation,
\be\label{firstlaw}
dE=T dS \qquad \longleftrightarrow \qquad dM=\frac{\kappa}{8\pi G}dA\,,
\ee
where $M$ is the black hole mass, $A$ is the area of the horizon and $\kappa$ is its surface gravity. Requiring the Euclidean solution to be regular at the horizon, one can further identify
\be\label{bhent}
T=\frac{\kappa}{2\pi}\,,\qquad\qquad S=\frac{A}{4G}\,,
\ee
as the black hole temperature and black hole entropy, respectively. Remarkably, Ryu and Takayanagi \cite{Ryu:2006bv} proposed that entanglement entropy $S_{EE}$, a measure of the entanglement between two subsystems of a general quantum system, can be computed holographically by
\be\label{ryu}
S_{EE}=\frac{\mathcal{A}}{4G}\,,
\ee
where $\mathcal{A}$ is the area of a certain extremal surface in the bulk. In addition to this striking similarity, it was later realized that entanglement entropy also satisfies a ``first law'' relation reminiscent of standard thermodynamical systems \cite{Blanco:2013joa}
\begin{equation}\label{1stLawEntanglement}
\delta S_{EE} =\delta\langle H_A\rangle\,.
\end{equation}
This equation relates the first-order variation of the entanglement entropy for a spatial region $A$ with the first-order variation of the expectation value of the ``modular Hamiltonian'' $H_A$, defined as the logarithm of the unperturbed reduced density matrix, $\rho_A\simeq e^{-H_A}$. Unfortunately, the modular Hamiltonian cannot always be expressed in terms of local operators. However, for spherical entangling regions in the vacuum of a conformal field theory (CFT) the modular Hamiltonian is given by a simple integral \cite{Casini:2011kv}
\be\label{modularH}
H_A=2\pi\int_A d^{d-1}x\frac{R^2-|\vec{x}-\vec{x}_0|^2}{2R}T_{00}\,,
\ee
where $T_{00}$ is the energy density of the CFT, $R$ is the sphere's radius and $\vec{x}_0$ denotes the center of the sphere. Thus, for arbitrary small perturbations over the CFT vacuum, the entanglement entropy of a sphere is given by
\be\label{firstlawI}
\delta S_{EE}=2\pi\int_Ad^{d-1}x\frac{R^2-|\vec{x}-\vec{x}_0|^2}{2R}\delta\langle T_{00}\rangle\,.
\ee
One might wonder if this equation has a dual interpretation in the gravity side of the correspondence. The answer to this question is surprising and rather remarkable: for CFTs with a holographic dual, the first law of entanglement entropy (\ref{firstlawI}) together with the Ryu-Takayanagi prescription (\ref{ryu}), automatically implies that the bulk geometry satisfies the Einstein field equations \cite{Lashkari:2013koa, Faulkner:2013ica}, linearized above pure AdS! More generally, for theories in which the entanglement entropy is computed by more general Wald functionals, one obtains the linearized field equations for the associated higher-derivative gravity dual.

More recently,
the effects of including  the cosmological constant as a thermodynamical variable were studied in \cite{Kastor:2009wy,Cvetic:2010jb,Dolan:2011xt,Dolan:2012jh,Cai:2013qga,Altamirano:2014tva,Kubiznak:2014zwa,Armas:2015qsv}.
   This program of varying the state as well as the couplings  has been dubbed  ``extended black hole thermodynamics'' or ``black hole chemistry'' since, in this context, the cosmological constant is associated with the pressure of the gravitational system, $P=-\Lambda/8\pi G$, while its conjugate quantity is identified as the thermodynamical volume $V$. We emphasize that, unlike parameters like  mass and charge that define the solution, $\Lambda$ also appears at the level of the action, so it is nondynamical. Nonetheless, it is still natural to ask how the laws of black holes thermodynamics are modified if we allow for such variations. For example, the first law is extended to
\be\label{firstlaw}
dE=TdS+VdP\,,
\ee
which is very simple and intuitive. In simple cases such as the Schwarzschild-AdS or the Reissner-Nordstr\"{o}m-AdS black hole in $(d+1)$ spacetime dimensions, the thermodynamical volume is shown to coincide with a naive integration over the black hole interior (in the Schwarzschild slicing),
\be\label{bhvolume}
V=\frac{\Omega_{d-1}r_+^{d}}{d}\,,
\ee
but its physical interpretation is still unclear.\footnote{In \cite{Armas:2015qsv} it was recently shown that black holes in AdS satisfy an infinite tower of extended first laws depending on which power of $\Lambda$ is varied, each of these with a different conjugate variable. This suggests that the formula (\ref{bhvolume}) for the black hole volume might not have any special physical meaning.} It is interesting to ask about the significance of this extended framework for gravitational theories with a holographic dual. As argued in \cite{Johnson:2014yja,Caceres:2015vsa}, in theories that arise as a consistent truncation of string/M theory, the value of the AdS radius $L$ is set by the value of the Planck length $\lP$, and the number of branes $N$. The worl dvolume theory is described in terms of a gauge theory with symmetries specified by the specific brane configuration; typically $N$ is the rank of the gauge group so it determines the number of degrees of freedom the theory. Newton's constant $G$ also depends nontrivially on $N$, so at the end one finds that
\begin{equation}\label{dictionary}
\frac{L^{d-1}}{G}\sim N^2\,.
\end{equation}
Thus, in this sense, varying the cosmological constant $\Lambda$ (and hence the $L$), is equivalent to changing the field theory to which the bulk background is dual. Furthermore, the conjugate variable associated to variations in $N$ can be interpreted holographically as a chemical potential for color \cite{Dolan:2014cja}. However, a careful application of the holographic dictionary teaches us that  varying $\Lambda$ also changes the volume of the field theory by changing the radius of curvature $R$ of the CFT metric \cite{Karch:2015rpa}.\footnote{If the field theory is defined on flat space, $R$ still sets the overall length scale of the theory, i.e. all the volumes scale as $V\sim R^{d-1}$.} In order to distinguish between these two effects, we observe \cite{Karch:2015rpa}  that for any function $f$  we have the dictionary:\footnote{In \cite{Karch:2015rpa}  this dictionary was used  to derive the generalized Smarr relation for AdS black holes from the scaling laws of CFT thermodynamics at large-$N$.}
\begin{equation}\label{karchdict}
\partial_{N^{2}}f |_{R} = \partial_{G^{-1}}f|_{L}\,,\qquad \partial_{R}f |_{N^{2}} = \partial_{L} f |_{L^{d-1}/G}\,.
\end{equation}
In other words, if we want to vary $N$ and keep $R$ fixed, we have to vary  Newton's constant $G$ in the bulk with the AdS length fixed; and if we want to vary $R$ and keep $N$ fixed, then we have to vary $L$ but keep the combination $L^{d-1}/G$ fixed.\footnote{In contrast, the relevant variation that appears in all other black hole chemistry literature, $\partial_L f|_G$, corresponds to changing both $N$ and $R$.} Now, these gravity couplings might or might not arise dynamically from a fundamental theory. For example, in the standard $D3$-brane system we can think of varying the number of branes $N$ and the Planck length $l_P$, which are non-dynamical. This, in the five-dimensional effective description corresponds to varying the cosmological constant and  Newton's constant. But there are also  examples in which field theory parameters arise dynamically from bulk fields. For example, in \cite{Caceres:2015vsa} the cosmological constant arises due to a scalar field that gets frozen to the minimum of its potential. Finally, we can also think of varying other couplings in the gravity side, for example higher derivative couplings, and in the boundary theory all these variations will also be associated with different field theory parameters. Thus, varying such couplings can be thought of as inducing a particular renormalization group (RG) flow in the spaces of theories.

One might wonder if there is an equivalent version of the extended first law of thermodynamics that applies for entanglement entropy and if so, what the dual interpretation might be. If so, this can be particularly useful to probe the structure of RG-fows as we explained above; ultimately, we would like to have
a better understanding of the phase transitions previously discovered in this context (e.g. the van der Waals transition for charged AdS black holes, see \cite{Kubiznak:2012wp}). The present paper is devoted to answering this question in the affirmative. In our approach, we make use of an extension of the Iyer-Wald formalism used in \cite{Faulkner:2013ica} to derive the first law of entanglement, but we include variations of both the cosmological constant $\Lambda$ and Newton's constant $G$. We perform our computation both in Einstein gravity and in higher-derivative theories, in which case we include additional field theory variations corresponding to the extra gravitational couplings. This study complements the existing approaches \cite{Kastor:2010gq, Kastor:2014dra} which rely on the Hamiltonian formulation of general relativity.\footnote{While our paper was in the final stage of preparation, the paper \cite{Kastor:2016bph} appeared, which contains some overlapping results.}

\subsection{Road map and summary}
The Iyer-Wald formalism is a powerful framework that provides a beautiful proof of the first law of black hole thermodynamics. The emphasis of the formalism on boundary terms and Stokes theorem makes it well suited for the holographic context, as it provides a means to translate between the bulk local language of differential geometry and the boundary non-local language of entanglement (and, more generally, quantum information theory). In the present work we make extensive use of this formalism in a different and more general context than the one associated with black hole thermodynamic. Thus, we would like to provide  an overview  of the present  work to  help the reader navigate  the next sections.

Consider a diffeomorphism invariant Lagrangian density ${\textbf L}$.
Let $\xi$ be  an arbitrary fixed vector in the $(d+1)$ spacetime under consideration. The variation of the Lagrangian under a diffeomorphism generated by $\xi^\mu$ is $\delta_\xi\textbf{L} = d (\xi \cdot \textbf{L})$. We can associate to $\xi$ a current
\begin{equation}
\textbf{J}=\bf{\Theta}(\delta_\xi\phi) - \xi\cdot\textbf{L}
\end{equation}
that will be conserved, $d\textbf{J}=0$, when the equations of motion are satisfied. If $\textbf{J}$ is conserved we can define the Noether  charge  $\textbf{Q}$ such that
$
\textbf{J}=d\textbf{Q}.
$
Now consider a  variation  $\delta\textbf{J}$
\begin{eqnarray}\label{eq:deltaJ}
\delta\textbf{J} & =& \delta{\bf \Theta}{(\phi,\pounds_\xi \phi)} -\xi\cdot \delta\textbf{L}\,, \\
&=& \delta {\bf \Theta}(\phi,\pounds_\xi \phi) - \pounds_\xi {\bf \Theta}(\phi,\delta\phi)+d (\xi\cdot{ \bf \Theta}(\phi,\delta\phi))\,.
\end{eqnarray}
If  we choose  $\xi$ such that it is a symmetry of all the fields, $\pounds_\xi \phi=0$ we have
\begin{equation}
\delta\textbf{J}  - d (\xi\cdot \bf{\Theta})=0\,.
\end{equation}
If in addition $\phi$ satisfies the equations of motion, we can replace $\delta{\textbf{J}}$ by $d\delta\textbf{Q}$ to obtain
\begin{equation}\label{eq:dchi}
d(\delta\textbf{Q}-\xi\cdot \bf{\Theta})=0\,.
\end{equation}
Integrating over a Cauchy surface of which the boundary is  $\partial \Sigma$,
\begin{equation}
\int_{\partial \Sigma} (\delta\textbf{Q}-\xi\cdot\bf{\Theta}(\phi,\delta\phi))=0\,.
\end{equation}
If we want to make contact with black hole thermodynamics we choose $\xi$ to be the time-like Killing vector that is null at the horizon and $\Sigma $ the corresponding  bifurcating surface. $\partial\Sigma$  will have two components, one at infinity and one at the horizon. The first law then follows from
\begin{equation}\label{eq:IWthermo}
\int_{\partial \Sigma_\infty} (\delta\textbf{Q}-\xi\cdot{\bf\Theta})=\int_{\partial \Sigma_{horizon}} (\delta\textbf{Q}-\xi\cdot\bf{\Theta})\,.
\end{equation}
The integral at infinity is the variation in the canonical energy, $\delta E$, while  the integral at the horizon is  $\frac{\kappa}{2 \pi} \delta S$.

We can proceed in a similar way  to obtain a first law of entanglement instead of a first law of thermodynamics. For a spherical boundary region in AdS the corresponding Ryu-Takayanagi surface is a bifurcating surface of a Killing vector field. Thus,  we can integrate \eqref{eq:IWthermo} not over the boundary of space time and the horizon but over the Ryu-Takayanagi surface and the boundary region. In that case the righthand side of \eqref{eq:IWthermo} will yield the entanglement entropy. Note that for a black hole it is no longer true that the Ryu-Takayanagi surface is a bifurcating surface of a Killing vector field and how to derive a first law of entanglement for excited states is still an open question.

Motivated by the possible field theory implications,  we  generalize \eqref{eq:dchi} to include variations in the couplings of the theory and  obtain an {\em extended} first law of entanglement. We find,
\begin{equation}\label{masterequation_rm}
    \sum_{i} \int_{\Sigma} \xi \cdot \textbf{E}^{c_{i}} \delta c_{i} + \int_{\partial \Sigma}(\delta\textbf{Q}-\xi\cdot\bf{\Theta}(\phi,\delta\phi)) = 0\,,
\end{equation}
where  $c_i$ denotes $\Lambda,\, G$ and any other  coupling of the theory. Equation \eqref{masterequation_rm} is one of the results of this paper. Section \ref{SecII} contains a  derivation of this result.

After having established the framework needed, in Section \ref{SecIII} we apply it to Einstein gravity in $(d+1)$ dimensions and  derive a first law of entanglement with variable  cosmological constant $\Lambda$ (or equivalently variable $L$)  and variable  Newton's constant $G$. For the sake of clarity we analyze each perturbation ($\delta L, \delta G$ and $\delta g_{\mu \nu}$) separately and after having calculated them we consider their joint effect to obtain,
\begin{equation}\label{ExtendedFirstLawEinstein_rm}
\delta E = \delta S_{EE} - (d-1)S_{EE}\frac{\delta L}{L} + S_{EE} \frac{\delta G}{G}\,.
\end{equation}
As usual, $E$ in this case is interpreted as the energy associated to the time evolution under the modular Hamiltonian $H_A$. We observe that 
\eqref{ExtendedFirstLawEinstein_rm} can be rewritten in terms of the variation of the central charge $c$:
\begin{equation}
\delta E = \delta S_{EE} - \frac{S_{EE}}{c}\delta c\,.
\end{equation}

In Section \ref{SecIV} we consider Gauss-Bonnet gravity and  derive an extended first law of entanglement with variable $\Lambda$, $G$ and variable Gauss-Bonnet coupling $\alpha$. Our result is,
\begin{equation}
\delta E = \delta S_{EE} - S_{EE}(c_L \delta L - c_G \delta G - c_\alpha \delta \alpha)\,,
\end{equation}
where the $c_L, c_G$ and $c_\alpha$ are constant coefficients that involve $d, L, G$ and $\alpha$. A similar expression is obtained for Lovelock gravity.

We conclude with Section \ref{SecV} where we  elaborate on the field theory interpretation of our results and discuss several open questions and possible directions of research  related to our work.

\section{Extended entanglement thermodynamics}\label{SecII}
The language of thermodynamics provides a natural framework to describe quantum entanglement: the reduced density matrix of a sphere in a CFT vacuum is thermal in nature. This fact was central to an early proof of the Ryu-Takayanagi formula for spherical regions \cite{Casini:2011kv}. While thermodynamics deals with systems in equilibrium, quantum entanglement is a powerful tool to probe out-of-equilibrium systems. Thus, formulating entanglement physics with thermodynamics may ultimately help understand out-of-equilibrium physics better by bringing it to a more familiar setting.

As mentioned in the Introduction, the first law of entanglement (\ref{1stLawEntanglement}) makes no reference to a pressure-volume conjugate pair, and the question naturally arises as to whether one can identify such quantities in order to capture the entanglement pattern of the state in a meaningful way. Several approaches exist in the literature to address this question. For example, \cite{Allahbakhshi:2013rda} suggests defining the entanglement pressure as the expectation value of the (spatial components of the) stress-energy tensor. In equilibrium, the entanglement pressure according to this approach would reduce to the field theory pressure. Another example is Jacobson's recent work \cite{Jacobson:2015hqa}, where a first law is studied with the variations of both the CFT state and the geometry. This yields an intriguing notion of pressure of which the  microscopic significance deserves further study.

In this work, we suggest taking the viewpoint of the black hole chemistry program, and identifying the pressure as the cosmological constant in the first law of entanglement. Let us consider the superposition of two perturbations in the bulk: the usual normalizable mode which is dual to perturbing the CFT state slightly away from the vacuum, and a perturbation of the cosmological constant. The combined effect of these two perturbations can be packaged into an extension of the first law:
\begin{equation}\label{Extended1stLawEE}
\delta \langle H_{A} \rangle = \delta S_{A} + V \delta P
\end{equation}
where the $\delta \langle H_{A} \rangle$ is due entirely to the normalizable mode in the bulk, but $\delta S_{A}$ is due to both perturbations. It follows that the volume is given by:
\begin{equation}\label{V}
    V = \frac{\partial S}{\partial P}
\end{equation}
If we believe the black hole chemistry program, the extended first law (\ref{Extended1stLawEE}) is quite natural: the CHM trick \cite{Casini:2011kv} can be used to map the first law of entanglement to the first law of black hole thermodynamics, therefore any meaningful notion of black hole volume seems meaningful to the entanglement first law. On the boundary side, varying the AdS length scale seems to correspond to some notion of changing the number of degrees of freedom in the field theory. For example, in 3 bulk dimensions the Brown-Henneaux says:
\begin{equation}\label{BrownHenneaux}
c = \frac{3L}{2G}
\end{equation}
Thus, changing $L$ (at fixed Newton's constant) amounts to varying the central charge of the CFT. have an RG-like flow in the space of theories, with potentially interesting structures.

Let us compute the entanglement volume for a sphere in a $d$-dimensional CFT vacuum with radius $R$. If the bulk is Einstein gravity, the Ryu-Takayanagi surface is a hemisphere in Poincar\'{e} coordinates $z^{2} + r^{2} = R^{2}$ and its area is the entanglement entropy:
\begin{equation}\label{SEEEinstein}
    S_{EE} = \frac{RL^{d-1}}{4G} \Omega_{d-2} \int_{0}^{\sqrt{R^{2}-\epsilon^{2}}} \frac{r^{d-2}}{(R^{2}-r^{2})^{d/2}} dr
\end{equation}
where $\Omega_{d-2}$ is the volume of a unit $(d-2)$-sphere, and we cut off the surface as usual at $z=\epsilon$. In this case the pressure dependence (or equivalently $L$ dependence) is quite trivial: it is simply an overall factor which is a power of $L$. We find the volume to be:
\begin{equation}
V = -\left( \frac{d-1}{2} \right) \frac{S}{P}
\end{equation}
While this result seems quite trivial, we stress that it is specific to Einstein gravity, and reflects the fact that in Einstein gravity the Ads length scale is essentially equivalent to the cosmological constant, which is a coupling in the theory. This special feature is lost in higher derivative theories such as Lovelock theories (which will be considered later in this paper): in such theories the AdS length scale is a complicated function of the couplings appearing in the gravity action, and varying these couplings yields a much richer structure.

Like the holographic dictionary (\ref{dictionary}) mentioned in the introduction, the Brown-Henneaux formula (\ref{BrownHenneaux}) implies that a variation of Newton's constant at fixed $L$ also results in varying the central charge of the CFT. In fact, it is argued in \cite{Karch:2015rpa} that this is perhaps the preferred way to vary the central charge, because a variation of $L$ at fixed $G$ actually also changes length scales in the CFT. We will come back to this issue and discuss it in greater details in the conclusion (section \ref{SecV}). This observation, however, motivates us to include in the extended first law the variation of $G$, as well as any other couplings appearing in the gravity action. The critical reader might object that, unlike the cosmological constant which plays the role of the pressure, there are not really any thermodynamic interpretations for the other couplings. There are precedents for this in the black hole literature, however. For example, the standard first law for Kaluza-Klein black holes includes the variation of the compactification radius, the thermodynamic conjugate of which is interpreted as a tension \cite{Kastor:2007wr, Kastor:2006ti}. An even more critical reader may also object that there is as of now no microscopic understanding of the cosmological constant as a pressure variable, in contrast with the entropy or temperature variables which can be obtained from the path integral. While this is true, we emphasize that the nature of the cosmological constant is far from settled, and that it is important to keep an open mind. At least from the gravity viewpoint, it naturally plays the role of a pressure like quantity.

In this paper, we will leave aside the hard questions of what the nature or structure of this flow in the space of theories (obtained by varying the gravity couplings) is, or what the microscopic picture behind the pressure as the cosmological constant might be. Instead, we ask the question of whether existing techniques in general relativity to derive the standard first law can be adapted to accommodate variations the couplings. This certainly is an interesting, if somewhat technical, question (see \cite{Kastor:2010gq, Kastor:2016bph, Kastor:2009wy, Kastor:2014dra} for a sample of existing papers along this line). Existing techniques to derive black hole thermodynamics fall under two broad categories: the Euclidean approach and the Noether charge approach. In the Euclidean approach, we analytically continue the time coordinate to obtain a geometry with conical defect. The entropy then comes from the gravitational action localized at the tip of the cone. The Euclidean approach was key to the proof of the Ryu-Takayanagi formula \cite{Lewkowycz:2013nqa}, since it is powerful enough to work even when the Ryu-Takayanagi surface is not the bifurcation of a Killing horizon, and therefore cannot be mapped with the CHM trick to a black hole horizon. On the other hand, the Noether charge approach (or the Iyer-Wald formalism \cite{Wald:1993nt, Iyer:1994ys}) is more restrictive: it demands a bifurcate Killing horizon ! On the other hand, it yields deeper insights into the nature of the entropy (namely, that it is intimately related to the diffeomorphism invariance of the theory). In the context of holography, the Iyer-Wald formalism has been instrumental in translating bulk geometrical quantities into quantum-information-theoretical quantities on the boundary \cite{VanRaamsdonk:2009ar, VanRaamsdonk:2010pw, Lashkari:2015hha}. For these reasons, we will apply the Iyer-Wald formalism in this paper. We will focus on the entanglement entropy of a sphere in the vacuum in a variety of gravity theories. In all the cases considered in this paper, the bulk is the Poincar\'{e} patch and the entangling surface is the usual hemisphere. This is indeed the bifurcation surface of a Killing horizon, so we meet the technical demands of the Iyer-Wald formalism.

\subsection{Iyer-Wald with varying the couplings}\label{SecIII}
In this subsection, we describe the (slight) generalization of the Iyer-Wald formalism needed to handle the variation of the couplings. In order not to hinder the general discussion, we will sketch out here the main steps and relegate the more technical details to Appendix \ref{IyerWaldAppendix}. Recall that the usual Iyer-Wald formalism is an algorithm which yields the first law via the computation of a few differential forms.

Consider a theory of gravity in $(d+1)$ dimensions. We first compute the symplectic potential current $\bf{\Theta}$, which is the boundary term obtained by varying the action under an arbitrary variation. Next, we consider a variation induced by an arbitrary vector field $\xi$ (i.e. the variation is the Lie derivative along $\xi$). By diffeomorphism invariance, we can compute the Noether current $\textbf{J}$ and Noether charge $\textbf{Q}$ associated to the symmetry generated by $\xi$. After finding the Noether charge and current, we then consider yet another kind of perturbation: an arbitrary on shell one. We then construct the form $\chi$:
\begin{equation}
    \chi = \delta \textbf{Q} - \xi \cdot \bf{\Theta}
\end{equation}
where $\delta \textbf{Q}$ is the variation of the Noether charge under the on shell perturbation, and $\bf{\Theta}$ is evaluated on this on shell perturbation. Finally, we specialize to the case where $\xi$ is a bifurcate Killing vector field. Then $\chi$ can be shown to be closed:
\begin{equation}
d\chi = 0
\end{equation}
Integrating over a spatial slice $\Sigma$ between the bifurcation surface $\mathcal{H}$ and infinity then yields the first law (using Stokes' theorem):
\begin{equation}\label{1stLawfromIW}
\int_{\Sigma} \chi = \int_{\infty} \chi - \int_{\mathcal{H}} \chi = 0
\end{equation}
In the black hole case, the integral over the bifurcation surface yields the $T\delta S$ term in the first law and the integral at infinity yields the $\delta M$ term. In the entanglement case, the integral over the bifurcation surface yields the variation of entanglement entropy $\delta S_{EE}$ and the integral at infinity yields the variation of the modular Hamiltonian $\delta \langle H \rangle$.

Let us now start by varying a coupling $c$ in the gravity action. Then the form $\chi$ is no longer closed. Instead of (\ref{1stLawfromIW}), we now find:
\begin{equation}\label{Extended1stfromIW}
\int_{\partial \Sigma}\chi + \delta c \int_{\Sigma} \frac{\partial \mathcal{L}}{\partial c} \xi \cdot \varepsilon = 0
\end{equation}
where, in the second term on the left, $\frac{\partial \mathcal{L}}{\partial c}$ is the partial derivative of the Lagrangian (both gravity and matter) with the coupling in question, and $\varepsilon$ is the volume form. Equation (\ref{Extended1stfromIW}) is the central result of this paper. As we will see in the examples below, the volume (as well as the conjugate to any coupling) receives contributions from two of the terms appearing in (\ref{Extended1stfromIW}): the new term proportional to $\delta c$, and  also the integral of $\chi$ at infinity.

To see that the volume indeed arises in this (somewhat complicated) way, suppose the only perturbation is $\delta L$. Since this does not result in a perturbation of the modular Hamiltonian, we expect the first law to be:
\begin{equation}
    0 = \delta S + V\delta P
\end{equation}
On the other hand, the Iyer-Wald formalism is designed so that the $\delta S$ term always arises from the integral of $\chi$ at the bifurcation surface. This is because the restriction of the Noether charge on the bifurcation surface reduces to the surface binormal, and its integral yields the area of the horizon. Therefore, the $V \delta P$ term must arise from the two other terms in (\ref{Extended1stfromIW}).

\subsection{Application to Einstein gravity}\label{Einstein}
In this subsection, we apply the technique developed above to derive the extended first law of entanglement for Einstein gravity. The bulk geometry is the $(d+1)$-dimensional Poincar\'{e} patch:
\begin{equation}
    ds^{2} = \frac{L^{2}}{z^{2}} (-dt^{2} + d\vec{x}^{2} + dz^{2})\,,
\end{equation}
where $\vec{x} = (x^{1}, x^{2}, \dots, x^{d-1})$. As previously mentioned, for a spherical boundary region with radius $R$, the Ryu-Takayanagi surface $ z^{2} + \vec{x}^{2} = R^{2}$ is the bifurcation surface of a Killing vector field $\xi$ gieven by:
\begin{equation}\label{KillingVectorField}
    \xi= -\frac{2 \pi}{R} t (z \partial_{z} + x^{i} \partial_{i}) + \frac{\pi}{R} (R^{2}-z^{2}-t^{2}-x^{2})\partial_{t}\,.
\end{equation}
Since we are only considering first order perturbations, we can turn them on one after another and in the end add up the results. First, the ordinary first law of entanglement can be obtained by turning on a normalizable mode in the bulk, resulting in a slightly excited state on the boundary. Since this part of the story is already well-known in the literature, we will not repeat it here but for the sake of completeness, we summarize the main steps in Appendix \ref{1stLawReview}.

Next, consider a perturbation of $L$. The perturbed metric takes the form:
\begin{equation}
    ds^{2} = \frac{L^{2}+2L\delta L}{z^{2}} (-dt^{2}+d\vec{x}^{2}+dz^{2})\,.
\end{equation}
The extended first law reads \eqref{masterequation}:
\begin{equation}\label{extended1stlawdeltaLEinstein}
    d (d-1)\frac{\delta( 1/L^2)}{16\pi G} \int_{\Sigma} \xi \cdot \varepsilon - \int_{\partial \Sigma_{\infty}} \chi + \int_{\partial \Sigma_{h}} \chi = 0\,.
\end{equation}
To compute the resulting perturbation of $\textbf{Q}$, we can simply compute the (unperturbed) Noether charge $\textbf{Q}$ and differentiate with respect to $L$. In Einstein gravity, the formula for the Noether charge is given in the Appendix (see equation (\ref{QinEinsteinHilbert})). Specializing to the AdS background and the Killing vector field above, we find the unperturbed Noether charge (restricted to the surface $t=0$, which contains the bifurcation surface) to be:
\begin{equation}\label{QonSigma}
    \textbf{Q}|_{\Sigma} = -\frac{1}{16\pi G} \left(\frac{4\pi z^{2} x^{i}}{RL^{2}} \varepsilon_{ti} + \frac{2z^{2}}{L^{2}} \left(\frac{2\pi z}{R} + \frac{\xi^{t}{(t=0)}}{z}\right) \varepsilon_{tz} \right)\,.
\end{equation}
In order to use the Iyer-Wald formalism, we need to calculate $\delta \textbf{Q}$ due to the shift in $L$. To do this, we can go back to the expression for the unperturbed Noether charge $\textbf{Q}$ (\ref{QonSigma}) and isolate the $L$ dependence of this expression. Notice that  $\varepsilon_{ti}$ and $\varepsilon_{tz}$ contain a factor of $\sqrt{-g}$, which is $(L/z)^{d+1}$. Therefore, the unperturbed Noether charge $\textbf{Q}$ depends on $L$ only through an overall factor of $L^{d-1}$. It follows that
\begin{equation}
    \delta \textbf{Q} = \frac{d-1}{L} \textbf{Q} \delta L\,.
\end{equation}
As for the symplectic potential current, we find that it vanishes (see Appendix \ref{app:theta0} for the details):
\begin{equation}\label{eqn325}
    {\bf{\Theta}} = 0\,.
\end{equation}
Therefore the Iyer-Wald form $\chi$ coincides with $\delta \textbf{Q}$. To extract the term with $\delta P$, we have to compute the integral of $\chi$ at infinity and the new term in the extended first law (as previously explained, the integral of $\chi$ over the horizon always gives the $\delta S$ term in the first law).

The restriction of $\chi$ to a cutoff $z=\epsilon$ near the boundary is:
\begin{equation}
    \chi |_{\partial \Sigma_{\infty}} = -\frac{(d-1)\delta L}{8GR} L^{d-2} \left(\frac{1}{\epsilon^{d-2}} + \frac{R^{2}-\vec{x}^{2}}{\epsilon^{d}} \right) dx^{1} \wedge \dots \wedge dx^{d-1}\,.
\end{equation}
Integrating over the boundary yields the divergent expression:
\begin{equation}\label{eq:Lpert_term2}
    \int_{\partial \Sigma_{\infty}} \chi = -\frac{(d-1)\delta L}{8GR}L^{d-2} \Omega_{d-2} \int_{0}^{\sqrt{R^{2}-\epsilon^{2}}} \left(\frac{1}{\epsilon^{d-2}} + \frac{R^{2}-r^{2}}{\epsilon^{d}} \right) r^{d-2}dr\,.
\end{equation}
Finally, evaluating the first term in  (\ref{extended1stlawdeltaLEinstein}), we find:
\begin{equation}
    d (d-1)\frac{\delta( 1/L^2)}{16\pi G} \int_{\Sigma} \xi \cdot \varepsilon = \frac{d(d-1)\delta L}{8GRL^{3}} \int_{\Sigma} (R^{2}-\vec{x}^{2}-z^{2})\left(\frac{L}{z}\right)^{d+1} dz \wedge dx^{1} \wedge \dots \wedge dx^{d-1}\,.
\end{equation}
If we perform the integral over $z$ in the expression above, we find:
\begin{eqnarray}\label{eq:Lpert_term3}
d (d-1)\frac{\delta( 1/L^2)}{16\pi G} \int_{\Sigma} \xi \cdot \varepsilon &=& -\frac{(d-1)\delta L}{8GR}L^{d-2} \Omega_{d-2} \int_{0}^{\sqrt{R^{2}-\epsilon^{2}}} \bigg(\frac{R^{2}-r^{2}}{\epsilon^{d}} + \frac{d}{2-d}\frac{1}{\epsilon^{d-2}} \nonumber \\
&+& \frac{2}{(d-2)}\frac{1}{(R^{2}-r^{2})^{\frac{d}{2}-1}} \bigg) r^{d-2}dr\,.
\end{eqnarray}
If we add \eqref{eq:Lpert_term2} and \eqref{eq:Lpert_term3},  we find that interestingly the divergences at order $\epsilon^{-d}$ in the integrands cancel, leaving us with:
\begin{equation}
    d (d-1)\frac{\delta( 1/L^2)}{16\pi G} \int_{\Sigma} \xi \cdot \varepsilon - \int_{\partial \Sigma_{\infty}} \chi  = \frac{(d-1)}{(d-2)}\frac{\delta L}{4GR}L^{d-2} \Omega_{d-2} \int_{0}^{\sqrt{R^{2}-\epsilon^{2}}} \left( \frac{d-1}{\epsilon^{d-2}} - \frac{1}{(R^{2}-r^{2})^{\frac{d}{2}-1}} \right) r^{d-2} dr\,.
\end{equation}
It can be shown that\footnote{This identity can be checked in a straightforward manner by noting that:
$$
    \int \frac{r^{d-2}}{(R^{2}-r^{2})^{\frac{d}{2}-1}} dr = \frac{r^{d-1}}{(R^{2}-r^{2})^{\frac{d}{2}-1}} - \frac{(d-2)}{(d-1)}\frac{r^{d-1}}{R^{d-2}} {}_{2}F_{1}{\left(\frac{d-1}{2},\frac{d}{2},\frac{d+1}{2},\frac{r^{2}}{R^{2}}\right)}
$$
and
$$
\int \frac{r^{d-2}}{(R^{2}-r^{2})^{d/2}}dr = \frac{1}{(d-1)}\frac{r^{d-1}}{R^{d}} {}_{2}F_{1}{\left(\frac{d-1}{2},\frac{d}{2},\frac{d+1}{2},\frac{r^{2}}{R^{2}} \right)\,.}.
$$}
\begin{equation}\label{integralidentity}
    \int_{0}^{\sqrt{R^{2}-\epsilon^{2}}} \left(\frac{d-1}{\epsilon^{d-2}} - \frac{1}{(R^{2}-r^{2})^{\frac{d}{2}-1}} \right) r^{d-2} dr = (d-2)R^{2} \int_{0}^{\sqrt{R^{2}-\epsilon^{2}}} \frac{r^{d-2}}{(R^{2}-r^{2})^{d/2}} dr\,.
\end{equation}
By comparing the right-hand side above with the integral (\ref{SEEEinstein}) for entanglement entropy, one easily recognizes that the two quantities are proportional to each other, and we find the first law:
\begin{equation}\label{eq:L_first}
    \delta^{(L)} S_{E} = \Psi_{L} \delta L\,,
\end{equation}
with the conjugate to $L$ given by
\begin{equation}
\Psi_{L} = (d-1)\frac{S_{EE}}{L}\,.
\end{equation}
The superscript $(L)$ in $\delta^{(L)}S_{E}$ is to emphasize that this is the contribution to $\delta S$ coming from a variation of $L$. Upon a trivial application of the chain rule, one can of course convert $\Psi_{L}$ to the volume $V$ defined in (\ref{V}).

Let us next consider a perturbation in $G$. Since the AdS metric does not explicitly depend on $G$, the metric perturbation vanishes:
\begin{equation}
\delta g_{\mu\nu} = 0\,.
\end{equation}
The extended first law in this case takes the form:
\begin{equation}\label{ExtendedfirstlawdeltaG}
-\frac{\delta G}{16\pi G^{2}} \int_{\Sigma} (R-2\Lambda) \xi \cdot \varepsilon - \int_{\partial \Sigma_{\infty}} \chi + \int_{\partial \Sigma_{h}} \chi = 0\,.
\end{equation}
Consider the variation of the Noether charge $\textbf{Q}$ under this perturbation. Since the unperturbed $\textbf{Q}$ depends on $G$ only through an overall factor of $G^{-1}$, we easily find:
\begin{equation}
\delta \textbf{Q} = -\frac{\delta G}{G}\textbf{Q}\,.
\end{equation}
Also, since the metric perturbation vanishes, the symplectic potential current trivially vanishes:
\begin{equation}
{\bf{\Theta}} = 0\,.
\end{equation}
Therefore, the Iyer-Wald form $\chi$ again coincides with $\delta \textbf{Q}$. Following the same steps as for the $\delta L$ perturbation, the restriction of $\chi$ to the boundary is:
\begin{equation}
\chi|_{\partial \Sigma_{\infty}} = \frac{L^{d-1}}{8G^{2}R}\delta G \left(\frac{1}{\epsilon^{d-2}} + \frac{R^{2}-\vec{x}^{2}}{\epsilon^{d}} \right) dx^{1} \wedge \dots \wedge dx^{d-1}\,.
\end{equation}
Integrating over the boundary, we find:
\begin{equation}\label{intchiboundarydeltaG}
    \int_{\partial \Sigma_{\infty}} \chi = \frac{L^{d-1}\delta G}{8G^{2}R} \Omega_{d-2} \int_{0}^{\sqrt{R^{2}-\epsilon^{2}}} \left( \frac{1}{\epsilon^{d-2}} + \frac{R^{2}-r^{2}}{\epsilon^{d}} \right) r^{d-2} dr\,.
\end{equation}
Finally, we evaluate the first term in  (\ref{ExtendedfirstlawdeltaG}). Since AdS$_{d+1}$ is maximally symmetric, we have:
\begin{equation}
R - 2\Lambda = -\frac{2d}{L^{2}}\,.
\end{equation}
We then find:
\begin{equation}
-\frac{\delta G}{16\pi G^{2}} \int_{\Sigma} (R-2\Lambda) \xi \cdot \varepsilon = \frac{d\delta G}{8G^{2}L^{2}R} \Omega_{d-2} \int_{\Sigma} (R^{2}-r^{2}-z^{2}) \left(\frac{L}{z}\right)^{d+1} r^{d-2} dz dr\,.
\end{equation}
As previously, we will explicitly do the integral over $z$ (from $\epsilon$ to $\sqrt{R^{2}-r^{2}}$),  yielding
\begin{equation}\label{eq:Gpert_term3}
-\frac{\delta G}{16\pi G^{2}} \int_{\Sigma} (R-2\Lambda) \xi \cdot \varepsilon = \frac{L^{d-1}\delta G}{8 G^{2}R} \Omega_{d-2} \int_{0}^{\sqrt{R^{2}-r^{2}}} \bigg(\frac{R^{2}-r^{2}}{\epsilon^{d}} + \frac{d}{2-d}\frac{1}{\epsilon^{d-2}}
+ \frac{2}{d-2}(R^{2}-r^{2})^{1-\frac{d}{2}} \bigg) r^{d-2} dr\,.
\end{equation}
Adding \eqref{intchiboundarydeltaG} and \eqref{eq:Gpert_term3}  we find again that the leading divergences inside the two integrands cancel each other, and upon using the identity (\ref{integralidentity}) we finally obtain:
\begin{align}
-\frac{\delta G}{16\pi G^{2}} \int_{\Sigma} (R-2\Lambda) \xi \cdot \varepsilon - \int_{\partial \Sigma_{\infty}} \chi =\
\Psi_{G}\delta G \,,
\end{align}
with
\begin{equation}
\Psi_{G} = -\frac{S_{EE}}{G}\,.
\end{equation}
Finally, the first law with variable G reads,
\begin{equation}\label{eq:G_first_final}
\delta^{(G)} S_{EE}=  -\frac{S_{EE}}{G}
 \delta G\,.
\end{equation}
Superposing all the perturbations, we find all in all the extended first law for Einstein gravity:
\begin{equation}\label{ExtendedFirstLawEinstein}
\delta E = \delta S_{EE} - (d-1)S_{EE}\frac{\delta L}{L} + S_{EE} \frac{\delta G}{G}\,.
\end{equation}
In particular, for AdS$_{3}$, this simplifies to:
\begin{equation}
\delta E = \delta S_{EE} - S_{EE} \left(\frac{\delta L}{L} - \frac{\delta G}{G}\right)\,.
\end{equation}
By the Brown-Henneaux formula, this in turn can be rewritten in terms of the variation of the central charge $c$:
\begin{equation}
\delta E = \delta S_{EE} - \frac{S_{EE}}{c}\delta c\,.
\end{equation}

\section{Entanglement chemistry in higher derivative theories}\label{SecIV}
In this section, we move on to the more interesting case of higher derivatives theories of gravity, starting with Gauss-Bonnet gravity. In the context of holographic entanglement entropy a thoroughly studied theory of this type is Gauss-Bonnet gravity \cite{deBoer:2011wk,Hung:2011xb,Bhattacharyya:2013jma,Dong:2013qoa,Camps:2013zua,Caceres:2015bkr}. Gauss-Bonnet gravity has, in addition to $\Lambda$ and $G$, one more coupling: the Gauss-Bonnet coupling $\alpha$. We will allow for variations of all three couplings. After deriving the extended first law for Gauss-Bonnet gravity, we will discuss generalizations to Lovelock gravity.

 The Lagrangian for Gauss-Bonnet gravity $(d+1)$ dimensions is \footnote{It is well known that when $d=3$, the Gauss-Bonnet term in the action (\ref{GBaction}) is topological and its integral over spacetime yields the Euler characteristic of the manifold. Thus we restrict ourselves to $d\ge 4$.}:
\begin{equation}\label{GBaction}
    \mathcal{L} = \left(\frac{R-2\Lambda}{16\pi G} + \alpha \mathcal{L}_{(2)} \right) \varepsilon\,,
\end{equation}
with
\begin{equation}
\mathcal{L}_{(2)} = R_{abcd}R^{abcd}-4R_{ab}R^{ab} + R^{2}\,,
\end{equation}
and $\alpha$ is the Gauss-Bonnet coupling. The equation of motion for the action above reads:
\begin{equation}
    R_{ab} - \frac{1}{2}g_{ab}\left(R - 2\Lambda + 16\pi G \alpha \mathcal{L}_{(2)} \right) + 32\pi G \alpha \mathcal{H}^{(2)}_{ab} = 0\,,
\end{equation}
with
\begin{equation}
    \mathcal{H}^{(2)}_{ab} = R_{aijk}R_{b}{}^{ijk} - 2R_{ac}R_{b}^{c} - 2R_{aibj}R^{ij} + RR_{ab}\,.
\end{equation}
We will need the symplectic potential current and Noether charge for this theory. Since these expressions are rather cumbersome, we list them in Appendix \ref{IyerWaldAppendix}. Like Einstein gravity, Gauss-Bonnet gravity above admits AdS$_{d+1}$ as a solution \cite{Hung:2011xb}:
\begin{equation}
    ds^{2} = \frac{L^{2}}{z^{2}}(dz^{2} - dt^{2} + d\vec{x}^{2})\,.
\end{equation}
The AdS length scale is now related to $\Lambda$, $G$ and $\alpha$ by:\footnote{There is a second $AdS_{d+1}$ solution with a different AdS length scale. However it contains ghosts and will be ignored in this paper.}
\begin{equation}\label{LambdaandL}
L^{2} = -\frac{d(d-1)}{4\Lambda} \left(1 + \sqrt{1+ \frac{(d-3)(d-2)}{d(d-1)}128\pi G\alpha \Lambda } \right),
\end{equation}
or equivalently
\begin{equation}\label{LandLambda}
\Lambda = \frac{d(d-1)}{2L^{4}} \left(16\pi G\alpha (d-2)(d-3) - L^{2}\right)\,.
\end{equation}
For $\alpha = 0$, we recover Einstein gravity and the usual relation $\Lambda = -\frac{d(d-1)}{2L^{2}}$.

The Ryu-Takayanagi surface $\Sigma$ is no longer a minimal area, but is computed by the prescription of \cite{Jacobson:1993xs} according to which we have to minimize the following functional:\footnote{One would naively think that entanglement entropy in Einstein-Gauss-Bonnet is computed by the Wald entropy formula. However, as pointed out in \cite{Hung:2011xb}, the Wald entropy does not correctly reproduce CFT results. However, the Jacobson-Myers prescription only differs from the Wald entropy by terms involving the extrinsic curvature. In the case of a Killing horizon, such as here, such terms vanish and the two prescriptions agree.}
\begin{equation}\label{GBRTaction}
S = \frac{1}{4G} \int_{M} d^{d-1}x \sqrt{h} \left[1 + 32\pi G \alpha \mathcal{R} \right]\,,
\end{equation}
where $\mathcal{R}$ denotes the Ricci scalar of the induced metric on $M$.\footnote{We have omitted a Gibbons-Hawking term which is needed to make the variational problem well defined. Technically, the Gibbons-Hawking term contributions to the entanglement entropy but it only gives a UV term which drops out anyway when we consider the variation $\delta S$.} It can be checked that the Ryu-Takayanagi remains the hemisphere as in Einstein gravity, i.e.
\begin{equation}
z = \sqrt{R^{2}-\vec{x}^{2}}\,.
\end{equation}
In particular, the fact that the Ryu-Takayanagi surface is still the bifurcation sphere of a Killing vector field means we can apply the Iyer-Wald formalism. If we regularize entanglement entropy by a cutoff at $z = \epsilon$ (or equivalently at $r = \sqrt{R^{2}-\epsilon^{2}}$), then entanglement entropy in $d$ dimensions is given by the integral:
\begin{eqnarray}\label{SEEGBunperturbed}
    S = \frac{RL^{d-1}}{4G} \left(1 - (d-1)(d-2)\frac{32\pi G \alpha}{L^{2}}\right) \Omega_{d-2} \int_{0}^{\sqrt{R^{2}-\epsilon^{2}}} \frac{r^{d-2}}{(R^{2}-r^{2})^{d/2}} dr\,.
\end{eqnarray}
Note that, like in Einstein gravity, the entanglement entropy is proportional to the area of the same surface as in Einstein gravity.

To derive the first law, we note again that we can turn on each perturbation separately one after another. The usual first law of entanglement is due to a normalizable mode in the bulk. Since this part of the story is not the focus of this paper, we again relegate it to Appendix \ref{1stLawReview}.

To deal with the variations of the couplings, we will first need to evaluate the Noether charge and the symplectic potential current on the AdS$_{d+1}$ background, just like for Einstein gravity, then differentiate with the coupling of interest. While the expressions (\ref{ThetaGB}) and (\ref{QGB}) look very intimidating, we can take advantage of the fact that AdS is maximally symmetric, and the Riemann and Ricci tensors simplify considerably:
\begin{equation}\label{RiemannAdS}
R_{abcd} = -\frac{1}{L^{2}} (g_{ac}g_{bd}-g_{ad}g_{bc})\,,
\end{equation}
\begin{equation}\label{RicciAdS}
R_{ab} = -\frac{d}{L^{2}}g_{ab}\,,
\end{equation}
\begin{equation}\label{scalarAdS}
R = -\frac{d(d+1)}{L^{2}}\,.
\end{equation}
Substituting the formulas above into (\ref{QGB}) and (\ref{ThetaGB}), we find:
\begin{eqnarray}\label{QGBAdS}
\textbf{Q} &=& \left(-\frac{1}{16\pi G} + \frac{2\alpha}{L^{2}}(d-1)(d-2) \right) \nabla^{[a}\xi^{b]} \varepsilon_{ab},
\end{eqnarray}
\begin{equation}\label{ThetaGBAdS}
{\bf{\Theta}} = \varepsilon_{d} \left(\frac{1}{16\pi G} - \frac{2\alpha}{L^{2}}(d-1)(d-2) \right) \left(g^{df}\nabla^{e} \delta g_{ef} - g^{ef} \nabla^{d} \delta g_{ef} \right).
\end{equation}
Note the striking similarity with Einstein gravity: despite the complicated form of the symplectic potential current and Noether charge, when we evaluate them on a maximally symmetric background such as $AdS$, they become basically the same tensor as in Einstein gravity except for an overall factor. The overall factor is sensitive to the Gauss-Bonnet coupling and reduces to that of Einstein gravity when we set $\alpha=0$.

Let us now further specialize to the particular Killing vector field under consideration. The Noether charge then becomes:
\begin{equation}\label{UnperturbedQGB}
\textbf{Q} |_{\Sigma} = \left(-\frac{1}{16\pi G} + \frac{2\alpha}{L^{2}}(d-1)(d-2) \right) \left[ \frac{4\pi z^{2} x^{i}}{RL^{2}} \epsilon_{ti} + \frac{2z^{2}}{L^{2}} \left(\frac{2\pi z}{R} + \frac{\xi^{t}{(t=0)}}{z} \right)\varepsilon_{tz} \right].
\end{equation}

\subsection{Variation of $L$ and $G$}
We are now ready to vary the couplings. In the action we will think about $\Lambda$ as a function of $L$, $G$ and $\alpha$ as given in equation (\ref{LandLambda}):
\begin{equation}
\Lambda = \Lambda(L,G,\alpha)\,.
\end{equation}

starting with $L$ (at fixed $G$ and $\alpha$). The perturbed metric is:
\begin{equation}
    ds^{2} = \frac{L^{2}+2L\delta L}{z^{2}} (-dt^{2} + d\vec{x}^{2} + dz^{2})\,.
\end{equation}
The extended first law with $\delta L$ takes the form:
\begin{equation}\label{ExtendedfirstlawGBdeltaL}
\delta L \int_{\Sigma} \frac{\partial \mathcal{L}}{\partial L} \xi \cdot \varepsilon - \int_{\partial \Sigma_{\infty}} \chi + \int_{\partial \Sigma_{h}} \chi = 0\,.
\end{equation}
The perturbed Noether charge is easily obtained by differentiation of (\ref{UnperturbedQGB}):
\begin{equation}
\delta \textbf{Q} = (d-1)\left[-\frac{1}{16\pi G L^{3}} + \frac{2\alpha}{L^{5}}(d-2)(d-3)\right] \left[\frac{4\pi z^{2}x^{i}}{R}\varepsilon_{ti} + \frac{2\pi z}{R} (z^{2} + R^{2} - \vec{x}^{2}) \varepsilon_{tz} \right] \delta L\,.
\end{equation}
As for the symplectic potential current, it can be seen from equation (\ref{ThetaGBAdS}) that it is proportional to the symplectic potential current of Einstein gravity (obtained by turning off $\alpha$). We know from section \ref{SecIII} that the symplectic potential current vanishes in Einstein gravity under the perturbation $L \rightarrow L + \delta L$. Therefore it must also vanish in Gauss-Bonnet theory:
\begin{equation}
{\bf{\Theta}} = 0\,,
\end{equation}
and the Iyer-Wald form coincides with $\delta \textbf{Q}$. As in Einstein gravity, the Iyer-Wald formalism is designed so that the integral of $\chi$ over the bifurcation surface yields $\delta S$, and we should evaluate the two other terms in (\ref{ExtendedfirstlawGBdeltaL}) in order to obtain the conjugate to $L$. The steps involved are quite similar to the Einstein gravity case, so we will only show a few intermediate steps. For example, the restriction of $\chi$ to the boundary is:
\begin{equation}
\chi \bigg|_{\partial \Sigma_{\infty}} = \left[-\frac{1}{16\pi G} + \frac{2\alpha}{L^{2}}(d-2)(d-3) \right]\frac{2\pi(d-1)L^{d-2}}{R}\delta L \left(\frac{1}{\epsilon^{d-2}} +\frac{R^{2}-\vec{x}^{2}}{\epsilon^{d}} \right) dx^{1} \wedge \dots \wedge dx^{d-1}.
\end{equation}
To evaluate the first term in (\ref{ExtendedfirstlawGBdeltaL}), we have to keep in mind that the $L$ dependence is implicit inside $\Lambda$. The chain rule yields:
\begin{equation}
    \frac{\partial \mathcal{L}}{\partial L} = \frac{\partial \mathcal{L}}{\partial \Lambda} \frac{\partial \Lambda}{\partial L} = -\frac{d(d-1)}{8\pi G L^{3}} \left[ 1 - \frac{32\pi G \alpha (d-2)(d-3)}{L^{2}} \right]
\end{equation}
When we add up the two integrals giving rise to $\delta L$ term in the first law, a few things happen which are also very similar to the Einstein case: the $\epsilon^{d}$ divergence cancels between the two integrands, and using the same identity as in the Einstein case (equation (\ref{integralidentity})), we finally find
\begin{equation}
    \delta L \int_{\Sigma} \frac{\partial \mathcal{L}}{\partial L} \xi \cdot \varepsilon - \int_{\partial \Sigma_{\infty}} \chi  = \Psi_{L} \delta L \,,
\end{equation}
with the conjugate of $L$, denoted by $\Psi_{L}$, given by:
\begin{equation}\label{PsiLGB}
\Psi_{L} = \frac{(d-1)}{L}S_{EE} \left(\frac{L^{2}-32\pi G \alpha (d-2)(d-3)}{L^{2} - 32\pi G \alpha (d-1)(d-2)}\right)\,.
\end{equation}
The extended first law takes the form:
\begin{equation}
\delta^{(L)} S_{EE} = \Psi_{L} \delta L\,.
\end{equation}

Next, let us now vary $G$ at fixed $L$ and $\alpha$. Notice that the AdS metric does not depend on $G$ or $\alpha$, but only on $L$, which is fixed in this subsection. Therefore, the metric perturbation vanishes:
\begin{equation}\label{nodeltag}
    \delta g_{\mu\nu} = 0\,,
\end{equation}
The extended first law with $\delta G$ takes the form:
\begin{equation}\label{extendedfirstlawGBwithdeltaG}
\delta G \int_{\Sigma}\frac{\partial \mathcal{L}}{\partial G} \xi \cdot \varepsilon - \int_{\partial \Sigma_{\infty}} \chi + \int_{\partial \Sigma_{h}} \chi = 0\,,
\end{equation}
To find the variation of the Noether charge due to $\delta G$, we differentiate the unperturbed Noether charge (\ref{UnperturbedQGB}) with respect to $G$:
\begin{equation}
    \delta \textbf{Q}|_{\Sigma} = \frac{\delta G}{16\pi G^{2}} \left[\frac{4\pi z^{2}x^{i}}{RL^{2}}\varepsilon_{ti} + \frac{2z^{2}}{L^{2}} \left(\frac{2\pi z}{R} + \frac{\xi^{t}{(t=0)}}{z} \right)\varepsilon_{tz} \right].
\end{equation}
On the other hand, it follows trivially from the fact that there is no metric perturbation that the symplectic potential current vanishes:
\begin{equation}
{\bf{\Theta}} = 0\,.
\end{equation}
Therefore, the Iyer-Wald form $\chi$ coincides with $\delta \textbf{Q}$. The integral of $\chi$ over the horizon gives $\delta S$, of course, and we will compute the other two terms in (\ref{extendedfirstlawGBwithdeltaG}) to derive the conjugate of $G$. The restriction of $\chi$ to infinity is:
\begin{equation}\label{eq:2GBG}
\chi \bigg|_{\partial \Sigma_{\infty}} = \frac{L^{d-1}\delta G}{8 R G^{2}} \left( \frac{1}{\epsilon^{d-2}} + \frac{R^{2}-\vec{x}^{2}}{\epsilon^{d}} \right) dx^{1} \wedge \dots \wedge dx^{d-1}\,.
\end{equation}
Finally, in order to compute the first integral in  (\ref{extendedfirstlawGBwithdeltaG}), we differentiate the Lagrangian with respect to $G$. The Lagrangian depends on $G$ in two ways: there is an explicit overall dependence in the Einstein part, and an implicit dependence through $\Lambda$ (according to our choice of parametrization). We obtain
\begin{equation}
\frac{\partial \mathcal{L}}{\partial G} = -\frac{1}{16\pi G^{2}} \left[R - 2\Lambda + 2G\frac{\partial \Lambda}{\partial G} \right] = \frac{d}{8 \pi G^{2}L^{2}}\,,
\end{equation}
where in the second equality we used (\ref{LandLambda}). By combining the two integrals giving rise to the $\delta G$ term in the extended first law, we find again that the $\epsilon^{d}$ divergences in the integrands cancel and (with the help of identity (\ref{integralidentity})) we are left with,
\begin{equation}
\int_{\partial \Sigma_{\infty}} \chi - \delta G \int_{\Sigma} \frac{\partial \mathcal{L}}{\partial G} \xi \cdot \varepsilon\  = - \Psi_{G}\,\delta G\,,
\end{equation}
with
\begin{equation}\label{PsiGGB}
\Psi_{G} = \frac{S_{EE}}{G} \left(\frac{L^{2}}{L^{2}-(d-1)(d-2)32\pi G \alpha}\right),
\end{equation}
The extended first law with $\delta G$ reads:
\begin{equation}
\delta^{(G)} S_{EE} = \Psi_{G}\, \delta G\,.
\end{equation}

\subsection{Varying the Gauss-Bonnet coupling $\alpha$}
Finally, we derive the extended first law for entanglement with varying $\alpha$ and fixed $G$ and $L$. Since the metric is not explicitly dependent on $\alpha$, we have again:
\begin{equation}
\delta g_{\mu\nu} = 0\,.
\end{equation}
The extended first law with $\delta \alpha$ has the form:
\begin{equation}\label{extendedfirstlawdeltalalpha}
\delta \alpha \int_{\Sigma} \frac{\partial \mathcal{L}}{\partial \alpha} \xi \cdot \varepsilon - \int_{\partial \Sigma_{\infty}} \chi + \int_{\partial \Sigma_{h}} \chi = 0\,,
\end{equation}
Next, $\delta \textbf{Q}$ is found by differentiating the unperturbed Noether charge (\ref{UnperturbedQGB}) with respect to $\alpha$:
\begin{equation}
\delta \textbf{Q} = \frac{2(d-1)(d-2)}{L^{2}}\delta \alpha \left[ \frac{4\pi z^{2}x^{i}}{RL^{2}} \varepsilon_{ti} + \frac{2z^{2}}{L^{2}} \left(\frac{2\pi z}{R} + \frac{\xi^{t}{(t=0)}}{z} \right) \varepsilon_{tz} \right].
\end{equation}
Also, it follows from the fact that $\delta g$ vanishes that the symplectic potential current does also:
\begin{equation}
{\bf{\Theta}} = 0\,.
\end{equation}
Therefore, the Iyer-Wald form coincides with $\delta \textbf{Q}$. As usual, the integral of $\chi$ over the bifurcation surface yields $\delta S$, and we compute the other two integrals in (\ref{extendedfirstlawdeltalalpha}) to derive the conjugate to $\alpha$. The restriction of $\chi$ to the boundary is:
\begin{equation}
\chi \bigg|_{\partial \Sigma_{\infty}} = \frac{4\pi}{R}(d-1)(d-2)L^{d-3} \delta\alpha \left(\frac{1}{\epsilon^{d-2}} + \frac{R^{2}-\vec{x}^{2}}{\epsilon^{d}} \right) dx^{1} \wedge \dots \wedge dx^{d-1}\,.
\end{equation}
Finally, to evaluate the first integral in (\ref{extendedfirstlawdeltalalpha}), we differentiate the Lagrangian with respect to $\alpha$ (then evaluate on the AdS background) \footnote{To evaluate $\mathcal{L}_{(2)}$ for AdS$_{d+1}$, we used the formulae (\ref{RiemannAdS}), (\ref{RicciAdS}), and (\ref{scalarAdS}). The result is:
$$
\mathcal{L}_{(2)} = \frac{(d+1)d(d-1)(d-2)}{L^{4}}\,.
$$
}:
\begin{equation}
    \frac{\partial \mathcal{L}}{\partial \alpha} = - \frac{1}{8\pi G}\frac{\partial \Lambda}{\partial \alpha} + \mathcal{L}_{(2)} = \frac{4d(d-1)(d-2)}{L^{4}}\,.
\end{equation}
In the end, we find the statement
\begin{equation}
\delta \alpha \int_{\Sigma} \frac{\partial \mathcal{L}}{\partial \alpha} \xi \cdot \varepsilon - \int_{\partial \Sigma_{\infty}} \chi\  = \Psi_{\alpha}\delta \alpha \,,
\end{equation}
where
\begin{equation}\label{PsialphaGB}
\Psi_{\alpha} = -\frac{32\pi G (d-1)(d-2)}{L^{2}-(d-1)(d-2)32\pi G\alpha} S_{EE}\,,
\end{equation}
and the extended first law with $\delta \alpha$ reads:
\begin{equation}
\delta^{(\alpha)} S_{EE} = \Psi_{\alpha}\delta \alpha\,.
\end{equation}

\subsection{Extended first law of entanglement for Gauss-Bonnet}

We are ready now to write a general first law of entanglement for Gauss-Bonnet gravity where we allow for the AdS radius $L$, Newton's constant $G$ and the Gauss-Bonnet coupling $\alpha$ to be variable.
\begin{equation}
\delta E = \delta S_{EE} - \Psi_{L}\delta L - \Psi_{G}\delta G - \Psi_{\alpha}\delta \alpha\,.
\end{equation}
The conjugate quantities to $L$, $G$ and $\alpha$ are given in equations (\ref{PsiLGB}), (\ref{PsiGGB}) and (\ref{PsialphaGB}), respectively.
Note that $\Psi_L, \Psi_G$ and $\Psi_\alpha$ are all proportional to the entanglement entropy so we can write the first law as
\begin{equation}
\delta E = \delta S_{EE} - S_{EE}(c_L \delta L - c_G \delta G - c_\alpha \delta \alpha)\,,
\end{equation}
where the $c_L, c_G$ and $c_\alpha$ are constant coefficients that involve $d, L, G$ and $\alpha$.

We will elaborate on the implications of this extended first law for the dual field theory when we discuss the conclusions of this paper.

\subsection{Extension to Lovelock theories}
The Lagrangian density for Lovelock gravity is
\begin{equation}
\mathcal{L}=\sum_{m=0}^{[\frac{d+1}{2}]} \lambda_m \mathcal{L}_m\,,
\end{equation}
where
\begin{equation}
\mathcal{L}_m(g)=\frac{1}{2^m}\sqrt{-g} \delta^{a_1 b_1....a_m b_m}_{c_1 d_1...c_m d_m} R_ {a_1 b_1}^{\ \ \ \ \ c_1 d_1} ....R_{a_m b_m }^{\ \ \ \ \ c_m d_m}\,,
\end{equation}
with $\mathcal{L}_0= \sqrt{-g}$ and the generalized $\delta$ is defined as a product of  Kronecker delta functions or recursively,
\begin{equation}
\delta^{\alpha_1 \alpha_1....\alpha_m}_{\beta_1 \beta_2...\beta_m}=\sum_{i=1}^{m} (-1)^{i+1}\delta^{\alpha_1 }_{\beta_i}\delta^{\alpha_2 \alpha_3....\alpha_m}_{\beta_1..\hat\beta_i..\beta_m}\,.
\end{equation}
In a $(d+1)$-dimensional space, the maximum order of a Lovelock theory is $m_{\text{max}} =[(d+1)/2]$, where the brackets indicate the integer part of $\frac{d+1}{2}$. Note that $\mathcal{L}_1$ and  $\mathcal{L}_2$ yield the Einstein and  Gauss-Bonnet Lagrangians respectively. The first three couplings, in terms of $\Lambda$, $G$ and the Gauss-Bonnet coupling $\alpha$, are given by:
\begin{equation}
\lambda_{0} = -\frac{\Lambda}{8\pi G}\,,
\end{equation}
\begin{equation}
\lambda_{1} = \frac{1}{16\pi G}\,,
\end{equation}
\begin{equation}
\lambda_{2} = \alpha\,.
\end{equation}

The holographic entanglement entropy for a general higher-derivative theory is given by \cite{Dong:2013qoa,Camps:2013zua}
\begin{align}\label{eq:higher_general}
S_{EE}= 2\pi &\int d^{d-1} y \sqrt{g}\Big\{-\frac{\partial L}{\partial R_{\mu\rho\nu\sigma}}\epsilon_{\mu\rho}\epsilon_{\nu\sigma}+ \sum_\alpha \left(\frac{\partial^2 L}{\partial R_{\mu_1\rho_1 \nu_1 \sigma_1}\partial R_{\mu_2 \rho_2 \nu_2 \sigma_2}}\right)_\alpha \frac{2 K_{\lambda_1 \rho_1 \sigma_1 }K_{\lambda_2 \rho_2 \sigma_2}}{q_\alpha +1} \nonumber\\
& [(n_{\mu_1 \mu_2} n_{\nu_1 \nu_2} -\epsilon_{\mu_1 \mu_2 }\epsilon_{\nu_1 \nu_2})n^{\lambda_1 \lambda_2} + (n_{\mu_1 \mu_2} \epsilon_{\nu_1 \nu_2} +\epsilon_{\mu_1 \mu_2 }n_{\nu_1 \nu_2})\epsilon^{\lambda_1 \lambda_2}]\Big \}\,,
\end{align}
where $K_{\lambda \rho \sigma}$ is the extrinsic curvature of the co-dimension 2 surface, $\epsilon_{\mu \nu}$ and $n_{\mu \nu}$ are appropriately defined tensors.
Things simplify considerably if we consider Lovelock gravity. In this case the functional to minimize, \eqref{eq:higher_general}, becomes  \cite{Dong:2013qoa}
\begin{equation}\label{eq:EE_Lovelock_simp}
S_{EE}= -4\pi \sum_m^{[\frac{d+1}{2}]} m \lambda_m \int d^{d-1} x \sqrt{h} \mathcal{L}_{m-1}\, (h)\,,
\end{equation}
where $h$ is the induced metric on the codimension-2 surface.
A simple solution of a Lovelock theory is AdS space\footnote{For AdS to be a solution we need  at least one of the Lovelock couplings $\lambda_m$ to be  real and negative \cite{Kastor:2010gq}.}\begin{equation}
ds^2 = \frac{L^2}{z^2} \left( dz^2 -dt^2 + \sum_{i=1}^{d-1} dx_i^2 \right)\,,
\end{equation}
where $L$ is in general a function of all the Lovelock couplings $\lambda_{m}$ and the dimension $d$:
\begin{equation}
L = L{(G, \Lambda ,\lambda_{2}, \lambda_{3}, \dots, d)}\,.
\end{equation}
If we take the  boundary region to be a sphere, $\sum_i dx_i^2 = dr^2 + r^2 d\Omega_{d-2}^2$ , the induced metric is
\begin{equation}
h_{a b} dx^a dx^b = \frac{L^2}{z^2}[(\dot{ r}^2 + \dot{ z}^2) dv^2 +  r^2 d\Omega_{d-2}]\,,
\end{equation}
where $v$ parametrizes the minimal surface in the $(z,r)$ plane. In was shown in \cite{Hung:2011xb} that in this case the surface that minimizes \eqref{eq:EE_Lovelock_simp} is a  hemisphere,
\begin{equation}\label{eq:hemi}
r(v)= R \cos \left(\frac{v}{R} \right)\,, \quad \quad \qquad z(v)= R \sin \left( \frac{v}{R} \right)\,.
\end{equation}
Using \eqref{eq:hemi} to evaluate \eqref{eq:EE_Lovelock_simp}, we find that, even in Lovelock theory, the entanglement entropy is proportional to the area of the horizon (for a spherical entangling region on the boundary):
\begin{align}\label{eq:simple_love}
S_{EE}&= \left(\frac{1}{4G} + \sum_{i=2}^{[(d+1)/2]} f_{i}  \lambda_i \right) \int d^{d-1}x \sqrt{\textrm{det}h_{a b}}\,,\\
&= \left(\frac{L^{d-1}}{4G} + L^{d-1} \sum_{i=2}^{[(d+1)/2]} {f}_{i}  \lambda_i \right) \ \tilde{A}\,,
\end{align}
where $\tilde{A}$ is a dimensionless area which does not depend on any of the couplings, and $f_{i}$ is a collection of functions of all the couplings $\lambda_{j}$ as well as the dimension $d$:
\begin{equation}
f_{i} = f_{i}{(L,G,\lambda_{2},\lambda_{3}, \dots, d)}\,.
\end{equation}
In \cite{Hung:2011xb} it was shown that the prefactor in (\ref{eq:simple_love}) is proportional to the central charge $\propto a_d^*$, so it is easy to check that the first law extends to arbitrary Lovelock theories. In the Iyer-Wald formalism, let us describe schematically how such a simplification arises. To extract the boundary term ${\bf{\Theta}}$, we start by varying the Lovelock action:
\begin{equation}
\delta \mathcal{L}_p=\sum_{m=0}^{[\frac{d+1}{2}]} [ \lambda_m \delta \mathcal{L}_m + (\delta \lambda_m)  \mathcal{L}_m]\,,
\end{equation}
with
\begin{equation}\label{deltaLm}
\delta \mathcal{L}_m(g)=\frac{m}{2^m}\sqrt{-g} R_ {a_1 b_1}^{\ \ \ \ \ c_1 d_1} ....R_{a_m b_m }^{\ \ \ \ \ c_m d_m} \delta^{a b a_{2} b_{2}....a_m b_m}_{c d c_{1} d_{1}...c_m d_m} \delta R_{a b}^{\ \ \ \ c d} + \dots\,,
\end{equation}
where the ellipsis on the right-hand side is for the term with $\delta (\sqrt{-g})$; this term contributes to the equation of motion exclusively and not the boundary term, so we did not write it down. We can now evaluate each of the $R_{ab}^{\\\\cd}$ factors on the AdS background. Once again, the symmetries of AdS come to our rescue, since:
\begin{equation}
R_{ab}{}^{cd} = -\frac{1}{L^{2}} (\delta_{a}^{c} \delta_{b}^{d} - \delta_{a}^{d}\delta_{b}^{c})\,.
\end{equation}
Therefore, evaluating (\ref{deltaLm}) on the AdS background simplifies to a matter of contracting Kronecker deltas! Therefore, for a spherical region and empty AdS the calculation in  Lovelock gravity proceeds very similarly to the one in previous sections and the extended first law of entanglement entropy takes the general form
\begin{equation}\label{extLovelock}
\delta E = \delta S_{EE} - S_{EE}\left(c_L \delta L - c_G \delta G -\sum_{i=2}^{[(d+1)/2]}c_{\lambda_i} \delta \lambda_i\right)\,,
\end{equation}
for some functions $c_{L}$, $c_{G}$ and $c_{\lambda_{i}}$, each of which depends in general on all the couplings as well as the dimension $d$. In terms of the central charge $a_d^*$ these are given by:
\be\label{extLovelock2}
c_{L}=\frac{1}{a_d^*}\frac{\partial a_d^*}{\partial L}\,,\qquad c_{G}=\frac{1}{a_d^*}\frac{\partial a_d^*}{\partial G}\,,\qquad c_{\lambda_i}=\frac{1}{a_d^*}\frac{\partial a_d^*}{\partial \lambda_i}\,.
\ee

\section{Final remarks and future directions}\label{SecV}

In this paper, we have presented an application of the extended black hole thermodynamics program to the area of entanglement entropy for CFTs with a gravity dual. The main result of the present work is an extended first law of entanglement that can be written schematically as follows: 
\begin{equation}\label{finalEQ}
\delta S_{EE} = \delta E + S_{EE}\sum_i c_i \delta \alpha_i\,.
\end{equation}
The first part of this equation, $\delta S_{EE} = \delta E$, is the standard first law of entanglement that arises by considering small variations of the quantum state around the vacuum of a CFT. As shown in \cite{Lashkari:2013koa, Faulkner:2013ica} this piece encodes the gravity equations of motion linearized around AdS. The second part of (\ref{finalEQ}) represents variations of field theory parameters dual to couplings in the gravity side of the correspondence. Here, we are denoting collectively the variations in $L$, $G$, and all higher derivative couplings as $\delta\alpha_i$. These new terms contain information about the gravity theory, which might not be encoded in the equations of motion. Consider, for example, Gauss-Bonnet in $d=3$ dimensions. In this case $\mathcal{L}_{(2)}$ is topological so the equations of motion are exactly the same as in Einstein gravity. In contrast, varying the coupling $\alpha$ gives a nontrivial effect in the extended first law, since the corresponding $c_{\alpha}$ does not vanish. Thus,
the extension of the first law gives \emph{off-shell} information about the dual gravity theory. In particular, given a collection of functions $c_i$, it is in principle possible to retrieve the value of all gravity couplings in the bulk action, by considering the appropriate variations in the dual CFT.

It is important to emphasize the different interpretations of the first and second pieces in (\ref{finalEQ}) from the CFT perspective: the first part refers to the change of the entanglement entropy due to an infinitesimal change in the quantum state of a theory, while the second part gives the change of entanglement entropy due to a change of the theory itself, staying always in their corresponding ground states. Incidentally, the formula (\ref{finalEQ}) can be intimately related to the extended first law of thermodynamics for AdS black holes, where one considers variations of the black hole horizon due to variations of the cosmological constant and other gravity couplings. To see this, recall that Minkowski space $\mathbb{R}^{d-1,1}$ can be conformally mapped to the hyperboloid $\mathbb{H}^{d-1}\times\mathbb{R}$, where the vacuum of the CFT is now interpreted as a thermal state. In the gravity side, this map is equivalent to a bulk diffemorphism that transforms AdS space into a topological black hole. In particular, the RT surface corresponding to a spherical region is mapped to the horizon of the topological black hole \cite{Casini:2011kv} so the entanglement entropy is reinterpreted as thermal entropy. The extended first law (\ref{finalEQ}) can then be recovered by considering the black hole chemistry of the topological black hole.

Let us focus for a moment on the extended first law in Einstein gravity, with arbitrary variations of $L$ and $G$. To interpret the new terms let us recall a basic formula in the holographic dictionary, namely
\begin{equation}\label{HolographicDictionary}
    \alpha \frac{L^{d-1}}{16\pi G} = N^{p}\,.
\end{equation}
Here, the coefficient $\alpha$ and the power $p$ are theory dependent. For a gauge theory, such as $\mathcal{N} = 4$ SYM, the power is $p=2$.
From the equation above, it follows that a variation of $L$ at fixed $G$ (or $G$ at fixed $L$) is equivalent to a variation in $N$ on the field theory side. This is what is usually done in the black hole chemistry literature. However, varying $L$ comes with an undesired side effect: that of varying the scale $R$ of the boundary metric. In general, variations of $L$ and $G$ can be translated to variations of $N$ and $R$ according to (\ref{karchdict}).
With these observations in mind, we now take another look at the extended first law for Einstein gravity given in (\ref{ExtendedFirstLawEinstein}),
\begin{equation}
\delta S_{EE} = \delta E + (d-1)S_{EE}\frac{\delta L}{L} - S_{EE} \frac{\delta G}{G}\,.
\end{equation}
If we now keep $L$ fixed, then we can trade $\delta G$ for $\delta (N^{2})$:
\begin{equation}
\frac{\delta G}{G} = -\frac{\delta (N^{2})}{N^{2}}\,,
\end{equation}
and the extended first law takes the form:
\begin{equation}
\delta S_{EE} = \delta E - \mu \delta (N^{2})\,,
\end{equation}
with the chemical potential corresponding to $N^{2}$ given by:
\begin{equation}\label{chemicalpotential}
\mu =  -\frac{S_{EE}}{N^{2}}\,.
\end{equation}
By contrast, recent works in the area of the extended thermodynamics typically interpret the coefficient of the $\delta L$ term as the chemical potential for color \cite{Kastor:2014dra}, which coincidentally gives the same result as (\ref{chemicalpotential}) above.

In holographic CFTs with a higher-derivative gravity dual is perhaps better to express the result in terms of the central charges, instead of $N$. In $(1+1) $dimensions, all CFTs are characterized by only one central charge $c$. As we have shown, in this case, the extended first law can be conveniently written as
\begin{equation}
\delta S_{EE} = \delta E + \frac{S_{EE}}{c}\delta c\,.
\end{equation}
Similarly, in section \ref{SecIV} we showed that for the class of theories we considered (Lovelock), the variations with respect to all gravity couplings can be collected in just one term:
\begin{equation}\label{1stlawastar}
\delta S_{EE} = \delta E + \frac{S_{EE}}{a_d^*}\delta a_d^*\,,
\end{equation}
so that the functions $c_i$ in (\ref{finalEQ}) can all be written as
\be
c_{i}=\frac{1}{a_d^*}\frac{\partial a_d^*}{\partial \alpha_i}\,.
\ee
The constant $a_d^*$ is a central charge that exists in an arbitrary number of dimensions and reduces to the coefficient of the A-type trace anomaly in even dimensions \cite{Myers:2010xs,Myers:2010tj}. It also satisfies a version of the $c$-theorem: it is monotonous under RG flows and $(a_d^*)_{\text{UV}}\geq(a_d^*)_{\text{IR}}$. Thus, equation (\ref{finalEQ}) encodes different ways in which we can change the central charge $a_d^*$ (varying different field theory parameters) and their corresponding changes in entanglement entropy. A straightforward observation is that, if we stay in the ground state:
\be
\frac{\delta S_{EE}}{S_{EE}}=\frac{\delta a_d^*}{a_d^*}\,.
\ee
Therefore, the entanglement entropy is also monotonous under changes of $a_d^*$. Specific variations with respect to individual couplings $\alpha_i$ do not need to be monotonous: they depend on the monotonicity properties of $a_d^*$ with respect to $\alpha_i$ (in the range of parameters allowed for each $\alpha_i$). It would be very interesting to arrive at similar result for excited states, and interpret the known $PV$ phase transitions (e.g. the van der Waals transition for charged AdS black holes \cite{Kubiznak:2012wp,Caceres:2015vsa}) in terms of a $c$-like theorem.


There are some open questions related to our work that are worth exploring:
\begin{itemize}
  \item \emph{Shape dependence.} From the field theory perspective, it is not clear if one can obtain a simple
  expression for the first law for general entangling surfaces. The reason is that the modular Hamiltonian cannot
  be expressed in terms of an integral over one-point functions as in (\ref{modularH}), but it generally depends on
  nonlocal data. From the bulk perspective, the complication arises because in this case the RT surface is not
  generally the bifurcation surface of a Killing vector field. In addition, even for the class of higher derivative
  theories we consider in this paper (i.e. Gauss-Bonnet and Lovelock) the functional for computing entanglement entropy
  (\ref{eq:simple_love}) picks up extra anomalous corrections coming from the second term of (\ref{eq:higher_general}).
  \item \emph{General higher-derivative theories.}\, It would be interesting to consider other examples of higher-derivative
  theories that might lead to simple functionals for entanglement entropy and try to derive the equivalent to the
  extended first law of entanglement. Some examples one can consider are $f(R)$ theories, quasitopological gravity and conformal gravity.
  The question to ask here is whether varying those extra couplings beyond Lovelock will encode extra information
  in the gravity side, which may be potentially rewarding. It would also be interesting to test if in these situations the variation of all gravity couplings could be reorganized in terms of some central charge of the boundary theory as in (\ref{1stlawastar}), or if the functions $c_i$ can be independent of each other.
  \item \emph{Nonlinear corrections}.\,  Obtaining the full non-linear Einstein equations from entanglement entropy is still an important problem in the context of AdS/CFT. In general, the positivity of relative entropy constrains the sign of higher-order perturbations \cite{Blanco:2013joa} but is not enough to derive the corrections to the equations of motion. Some recent progress was achieved recently in \cite{Lashkari:2015hha,Beach:2016ocq}.
      For holographic CFT states near the vacuum, entanglement entropy can be expressed perturbatively as an expansion in the one-point functions of local operators dual to light bulk fields. Using the connection between quantum Fisher information and canonical energy, the authors derived a general formula for such an expansion up to second-order in the one-point functions, extending the first-order result given by the entanglement first law. Following the same spirit but applied to our context, it would be interesting to extend our results by considering nonlinear corrections to the gravity couplings and to explore their implications.
  \item \emph{Extended first law in field theory.}\, The derivation of the extended first law of entanglement entropy presented in this paper relies completely on AdS/CFT methods. It would be interesting to come up with a simple field theory example where, starting with a family of CFTs labeled by central charges $a$ and $c$, one can compute the entanglement entropy $S_{EE}(a,c)$ and obtain the associated extended first law. A natural question here is to ask about the universality of (\ref{1stlawastar}). Does it work for general theories, or is it a properties of holographic CFTs?
  \item \emph{String / M-theory realizations.}\, There are a number of works that explore the extended thermodynamics of systems of branes in string and M theory \cite{Dolan:2014cja,Belhaj:2015uwa,Chabab:2015ytz}. These works treat the number of branes as a dynamical variable and study the associated phase space. It would be interesting two extend these results in two ways $i)$ consider variations in the string coupling $g_s$, which would be the equivalent of varying the Newton's constant $G$ in the low-energy effective theory  and  $ii)$ consider the computation of entanglement entropy and  the extended first law in these setups. 
  \item \emph{$1/N$ corrections.}\, In the context of holography, the leading loop corrections to entanglement entropy are given by the bulk entanglement entropy between the two bulk regions separated by the RT surface \cite{Faulkner:2013ana}. In general, such corrections introduce new divergences that depend on the bulk UV cutoff $\Lambda_{\text{bulk}}$, but are expected to cancel by the renormalization of Newton's constant $G$ (see e.g. \cite{Solodukhin:2011gn} and the references therein). It would be interesting  to study the interplay of these corrections with the classical variation we consider in this paper $\delta G$.
  \item \emph{Extended first law for excited states.\footnote{We thank Ted Jacobson for discussion and suggestions on how to approach this issue.} }\, It would be desirable to derive a version of the extended first law of entanglement for variations of the quantum state around an arbitrary excited state (not necessarily the CFT vacuum), for example, around a thermal state  (previous work on excited states includes \cite{Bhattacharya:2012mi,Allahbakhshi:2013rda,Wong:2013gua}). In the context of holography, such study may shed light on the results of \cite{Caceres:2015vsa}, which showed that entanglement entropy can be used as an efficient order parameter to uncover the thermodynamic phase transitions associated to the extended $PV$ space. It would be interesting to understand the connection of such transitions with holographic RG flows and $c$-theorems.
  \item \emph{Relation with holographic complexity.}\,  Another quantity that generalizes the concept of thermodynamical volume to the context of entanglement entropy is the recently proposed holographic complexity, computed by the volume associated to the entanglement wedge \cite{Alishahiha:2015rta}. Very recently it was argued that this quantity also captures the behavior of the extended $PV$ space \cite{Momeni:2016qfv}. It would be interesting to investigate if there is a more direct connection between complexity and the extended first law of entanglement.
  \item \emph{Black hole chemistry from Iyer-Wald.}\,  The extended Iyer-Wald formalism provides an alternative method for computing  the thermodynamical volume of black holes in general diffeomorphism invariant theories of gravity, as an integral of the black hole \emph{exterior} rather than its \emph{interior}. Therefore, the method might be very useful for studying black hole chemistry in problematic cases such as in Taub-NUT-AdS/Taub-Bolt-AdS \cite{Johnson:2014xza,Johnson:2014pwa,Lee:2014tma} and Lifshitz spacetimes \cite{Gim:2014nba,Brenna:2015pqa}.
\end{itemize}

We hope to come back to some of these problems in the near future.

\begin{acknowledgments}
It is a pleasure to thank Jay Armas, Jan de Boer, Niels Obers and Brandon Robinson for discussions and comments on the manuscript. J.F.P. also thank the organizers and participants of the NORDITA workshop ``Black Holes and Emergent Spacetime'' for stimulating discussions on the subject. This research was supported by Mexico's National Council of Science and Technology  Grant No. CB-2014-01-238734, the National Science Foundation Grant No. PHY-1620610 and the Foundation for Fundamental Research on Matter which is part of the Netherlands Organization for Scientific Research.
\end{acknowledgments}
\appendix

\section{Iyer-Wald with varying the couplings: a closer look}\label{IyerWaldAppendix}

In this section we will review and extend the Iyer-Wald formalism \cite{Wald:1993nt, Iyer:1994ys}  to include  variations in all the couplings of the theory. A similar treatment can be found in \cite{Urano:2009xn} for variations with respect to the cosmological constant only.\footnote{The paper \cite{Urano:2009xn} applies the formalism to study physics in de Sitter space.} Since our  main interest is the holographic implications of these variations and  field theory quantities typically involve combinations of the gravity coupling constants we will develop a framework to include  variations with respect to all the couplings appearing in the gravity  theory.

Consider a theory of gravity with diffeomorphism invariance coupled to matter. The Lagrangian can be written as a (d+1)-form:
\begin{equation}
    \textbf{L}{(g, \phi, c_{i})} = \mathcal{L} \varepsilon = \mathcal{L}_{g}{(g,c_{i})} \varepsilon + \mathcal{L}_{m}(\phi, g, c_{i}) \varepsilon\,,
\end{equation}
where $\mathcal{L}_{g}$ is the gravitational Lagrangian, $\mathcal{L}_{m}$ is the matter Lagrangian, $\phi$ stands for any matter, $\varepsilon$ is the volume element\footnote{The volume element is given by:
$$
    \varepsilon = \sqrt{-g} dt \wedge dx^{1} \wedge \dots \wedge dx^{d}
$$
For later convenience, we will also define the $d$-form:
$$
    \varepsilon_{a} = \frac{1}{d!} \epsilon_{a b_{2} \dots b_{d+1}} dx^{b_{2}} \wedge \dots \wedge dx^{b_{d+1}}
$$
and the $(d-1)$-form:
$$
    \varepsilon_{ab} = \frac{1}{(d-1)!}\varepsilon_{ab c_{3} \dots c_{d+1}} dx^{c_{3}} \wedge \dots \wedge dx^{c_{d+1}}
$$
where $\epsilon$ is the Levi-Civita tensor, with the sign convention $\epsilon_{tzx^{1} \dots x^{d-1}} = +\sqrt{-g}$.} and $c_{i}$ are the couplings of the gravitational theory. The variation of the Lagrangian takes the form:
\begin{equation}\label{eq2}
    \delta \textbf{L} = \textbf{E}^{g} \delta g + \textbf{E}^{\phi} \delta \phi + d{\bf{\Theta}}_{g} (g,\delta g) + d{\bf{\Theta}}_{m} (g,\phi,\delta g,\delta \phi) + \sum_{i} \textbf{E}^{c_{i}} \delta c_{i}\,,
\end{equation}
where $\textbf{E}^{g}$ is the Einstein field equation, $\textbf{E}^{\phi}$ is the Euler-Lagrange equation for the matter, $\textbf{E}^{c_{i}}$ is given by
\begin{equation}
    \textbf{E}^{c_{i}} = \frac{\partial \mathcal{L}}{\partial c_{i}}\varepsilon\,,
\end{equation}
where and ${\bf{\Theta}}_{g}$,\  ${\bf{\Theta}}_{m}$  are  the boundary terms obtained  when the gravitational action and the matter action  are varied. We will  use ${\bf{\Theta}}$ for the sum of the two boundary terms and refer to ${\bf{\Theta}}$ as the symplectic potential current.

The Iyer-Wald formalism derives the first law of black hole thermodynamics by considering two different kinds of variations: (1) first consider a variation generated by a vector field, (2) and then an arbitrary variation induced by bulk fields. First, let $\xi^{\mu}$ be an arbitrary vector field, and consider the field variation generated by $\xi^{\mu}$: $\delta_{\xi} = \mathcal{L}_{\xi}$. The Noether current associated with the coordinate transformation generated by $\xi$ is:
\begin{equation}
    \textbf{J} = {\bf{\Theta}}{(g,\phi,\delta_{\xi}g, \delta_{\xi}\phi)} - \xi \cdot \textbf{L}\,.
\end{equation}
The ``dot product'' in the second term on the right-hand side means the contraction of the vector field with the first index of the form.\footnote{For example, for an n-form $F = \frac{1}{n!} F_{a_{1} a_{2} \dots a_{n}} dx^{a_{1}} \wedge dx^{a_{2}} \wedge \dots \wedge dx^{n}$, we have $\xi \cdot F = \frac{1}{(n-1)!}\xi^{b} F_{b a_{2} \dots a_{n}} dx^{a_{2}} \wedge \dots \wedge dx^{a_{n}}$.} This Noether current is a $d$-form. Naturally, $\textbf{J}$ splits into a gravity current $\textbf{J}_{g}$ and a matter current $\textbf{J}_{m}$. We will now check that this current is conserved on shell, even with varying couplings $c_{i}$. To do this, we compute the exterior derivative of $\textbf{J}$:
\begin{equation}\label{dJ}
    d\textbf{J} = d{\bf{\Theta}}{(g,\phi,\delta_{\xi}g, \delta_{\xi}\phi)} - d(\xi \cdot \textbf{L})\,.
\end{equation}
After some manipulation, this can be cast as:\footnote{We use Cartan's magic formula:
$$
    \mathcal{L}_{\xi}\textbf{L} = \xi \cdot d\textbf{L} + d(\xi \cdot \textbf{L})
$$
We also used the fact that $d\textbf{L}=0$ since $\textbf{L}$ is a top-dimensional form, and equation (\ref{eq2}).}
\begin{equation}
    d\textbf{J} = - \textbf{E}^{g}\mathcal{L}_{\xi}g -  \textbf{E}^{\phi} \mathcal{L}_{\xi}\phi - \sum_{i} \textbf{E}^{c_{i}}\mathcal{L}_{\xi} c_{i}\,.
\end{equation}
The first two terms on the right-hand side vanish on shell. And the last term trivially vanishes since the couplings $c_{i}$ have no spacetime dependence. Therefore, we conclude that $d\textbf{J} = 0$ on shell, and $\textbf{J}$ is (locally) the exterior derivative of a $(d-2)$-form $Q$:
\begin{equation}\label{eq9}
    \textbf{J} = d\textbf{Q}\,.
\end{equation}
$\textbf{Q}$ is the Noether charge associated with the symmetry generated by $\xi$. Next, consider a variation of $\textbf{J}$ under an arbitrary variation (not induced by a vector field). We have:
\begin{equation}
    \delta \textbf{J} = \delta {\bf{\Theta}}{(g,\phi,\delta_{\xi} g, \delta_{\xi} \phi)} - \xi \cdot \delta \textbf{L}\,.
\end{equation}
Note that in the above equation we do not vary $\xi$ (i.e. $\delta \xi = 0$) since we do not consider $\xi$ as a dynamical variable in this formalism. After some manipulations, we find:
\begin{equation}\label{eq12}
    \delta \textbf{J} = \delta {\bf{\Theta}}{(g,\phi,\delta_{\xi} g,\delta_{\xi} \phi)} - \mathcal{L}_{\xi}{\bf{\Theta}}{(g,\phi,\delta g,\delta \phi)} + d(\xi \cdot {\bf{\Theta}}) - \sum_{i} \xi \cdot \textbf{E}^{c_{i}} \delta c_{i}\,.
\end{equation}
At this stage, it is convenient to introduce the symplectic  current ${\bf{\Omega}}$, defined by:
\begin{equation}
    {\bf{\Omega}}{(\psi,\delta_{1}\psi,\delta_{2}\psi)} = \delta_{1}[{\bf{\Theta}}{(\psi,\delta_{2}\psi)}] - \delta_{2}[{\bf{\Theta}}{(\psi,\delta_{1}\psi)}]\,,
\end{equation}
where $\psi$ stands for all the dynamical variables including the metric, and $\delta_{1}$, $\delta_{2}$ are two arbitrary variations. We can then rewrite equation (\ref{eq12}) as:
\begin{equation}
    \delta \textbf{J} = {\bf{\Omega}}{(g,\delta g,\delta_{\xi} g)} + d(\xi \cdot {\bf{\Theta}}) - \sum_{i} \xi \cdot \textbf{E}^{c_{i}}\delta c_{i}\,.
\end{equation}

Up to now we have considered an arbitraty vector $\xi$. Let us now specialize to  a Killing vector field, $\pounds_\xi g =0$. In this case the symplectic current vanishes. Using equation (\ref{eq9}), we then find\footnote{We replace $\delta \textbf{J}$ by $d \delta \textbf{Q}$. This is only allowed when the perturbations $\delta g$ and $\delta \phi$ are on shell in the sense that they satisfy the linearized equation of motion. Since we are varying the couplings, the linearized equation of motion must include additional terms containing the variation of these couplings.}
\begin{equation}
    d(\delta \textbf{Q} - \xi \cdot {\bf{\Theta}}) + \sum_{i} \xi \cdot \textbf{E}^{c_{i}} \delta c_{i} = 0\,.
\end{equation}
We now integrate the equation above over a codimension-1 hypersurface $\Sigma$ and use Stoke's theorem:
\begin{equation}\label{masterequation}
    \sum_{i} \int_{\Sigma} \xi \cdot \textbf{E}^{c_{i}} \delta c_{i} + \int_{\partial \Sigma} \chi = 0\,,
\end{equation}
with $\chi$ defined to be the form
\begin{equation}\label{eq:chi_def}
    \chi = \delta \textbf{Q} - \xi \cdot {\bf{\Theta}}\,.
\end{equation}
Equation \eqref{masterequation} is one of the results of this paper. In the following sections  we will make use of it to derive an extended first law of entanglement entropy.

Let us now give the explicit expressions for the symplectic potential current ${\bf{\Theta}}_{g}$ and Noether charge $\textbf{Q}$ for Einstein gravity and Gauss-Bonnet gravity. For Einstein gravity, we have:
\begin{align}
{\bf{\Theta}}_{g}& = \frac{1}{16\pi G_{N}} g^{ac}g^{bd} \left(\nabla_{b} \delta g_{c d} - \nabla_{c} \delta g_{bd} \right) \varepsilon_{a}\,, \\
    \textbf{Q}& = -\frac{1}{16\pi G} \nabla^{a}\xi^{b} \varepsilon_{ab}\,.\label{QinEinsteinHilbert}
\end{align}

For Gauss-Bonnet gravity, the symplectic potential current $d$-form was given in \cite{Iyer:1994ys}:
\begin{eqnarray}\label{ThetaGB}
    {\bf{\Theta}} &=& \varepsilon_{d} \bigg[ \bigg( \frac{1}{16\pi G} + 2 \alpha R \bigg) g^{de}g^{fh} (\nabla_{f}\delta g_{eh}-\nabla_{e}\delta g_{fh}) \nonumber \\
    &+& \alpha \bigg(-2(\nabla^{e} R)g^{df}\delta g_{ef} + 4R^{de} (\nabla_{e} \delta g_{fh})g^{fh} + 4 R^{ef} (\nabla^{d}\delta g_{ef}) \nonumber \\
    &-& 8 R^{ef} (\nabla_{e}\delta g_{fh})g^{dh} - 4(\nabla^{e}R^{df})\delta g_{ef} + 4R^{defh}\nabla_{h}\delta g_{ef} \bigg) \bigg]\,,
\end{eqnarray}
and the Noether charge $(d-1)$-form is \cite{Iyer:1994ys}:
\begin{equation}\label{QGB}
\textbf{Q} = -\varepsilon_{de} \left(\frac{1}{16\pi G}\nabla^{d}\xi^{e} + 2\alpha (R\nabla^{d}\xi^{e} + 4\nabla^{[f}\xi^{d]}R^{e}_{f} + R^{defh}\nabla_{f}\xi_{h}) \right)\,.
\end{equation}
A note here is in order about the expressions above for Gauss-Bonnet theory. Technically, the results of \cite{Iyer:1994ys} assume zero cosmological constant. However, for both Einstein and Gauss-Bonnet gravity, the introduction of a cosmological constant does not modify the boundary term $\Theta$ when we vary the action, so $\Theta$ remains the same in AdS as in flat space. As for the Noether current $\textbf{J}$ and Noether charge $\textbf{Q}$, the reader can check that their off-shell definitions will be modified by the presence of the cosmological constant but on shell they are also the same.\footnote{We thank Robert Wald for explaining this point.}

\section{Entanglement first law and linearized bulk e.o.m.:   a review}\label{1stLawReview}
In this Appendix, we review the equivalence between the (unextended) first law of entanglement and the linearized equation of motion in the bulk, both in Einstein gravity and Gauss-Bonnet gravity. The Einstein gravity case has been treated in \cite{Faulkner:2013ica, Blanco:2013joa}, which we follow closely. Consider a generic perturbation of AdS:
\begin{equation}\label{FeffermanGraham}
    ds^{2} = \frac{L^{2}}{z^{2}} \left(-dt^{2} + d\vec{x}^{2} + dz^{2} + z^{d} H_{\mu\nu}{(z,x,t)} dx^{\mu}dx^{\nu} \right)\,,
\end{equation}
where $\mu$ and $ \nu$ are the boundary coordinates $t$ and $x^{i}$. We work in the radial gauge where $H_{zt} = H_{zx} = H_{zz} = 0$. In order for the perturbation to solve the linearized Einstein equation, $H_{\mu\nu}$ has to be traceless ($H_{\mu}^{\mu}=0$), divergence free  ($\partial_{\mu}H^{\mu\nu}=0$) and satisfies:
\begin{equation}\label{linearizedEOMAdS}
    \frac{1}{z^{4}}\partial_{z} (z^{4}\partial_{z} H_{\mu\nu}) + \partial^{2}H_{\mu\nu} = 0
\end{equation}
Substituting the perturbed metric (\ref{FeffermanGraham}) into the formula for $\chi$ in Einstein gravity and working to first order in $H_{\mu\nu}$, we find:
\begin{eqnarray}
    \chi |_{\Sigma } &=& \frac{z^{d}}{16\pi G} \bigg\{ \varepsilon^{t}{}_{z} \left[\left(\frac{2\pi z}{R} + \frac{d}{z}\xi^{t} + \xi^{t}\partial_{z} \right) H^{i}_{i}\right]\,, \nonumber \\
    &+& \varepsilon^{t}{}_{i} \left[\left(\frac{2\pi x^{i}}{R} + \xi^{t}\partial^{i} \right)H^{j}_{j} - \left(\frac{2\pi x^{j}}{R} + \xi^{t}\partial^{j} \right) H^{i}_{j} \right] \bigg\}\,.
\end{eqnarray}
The restrictions to the boundary at infinity and to the bifurcating surface are:
\begin{align}\label{EinsteinChi}
\chi |_{\partial \Sigma_{\infty}} &= -\frac{L^{d-1}d}{16GR} (R^{2}-\vec{x}^{2}) H_{i}^{i}{(z=0)} dx^{1} \wedge \dots \wedge dx^{d-1}\,,\\
\chi |_{\partial \Sigma_{h}}& = -\frac{L^{d-1}}{8GR} (R^{2}H^{i}_{i} - x^{i}x^{j}H_{ij}) dx^{1} \wedge \dots \wedge dx^{d-1}\,.
\end{align}
The Iyer-Wald formalism states that:
\begin{equation}
\int_{\Sigma_\infty}\chi =\int_{\Sigma_h} \chi\,.
\end{equation}
The equality between the two quantities above can be verified directly by integrating (\ref{EinsteinChi}) over the boundary and bifurcation surface. As argued in the main body of the paper, the integral over the horizon necessarily gives $\delta S$. One can also directly check that the integral over the boundary gives $\delta E$. Recall that the energy associated to a Killing vector field $\xi$ is given by
\begin{equation}
    E = \int d\Sigma^{\mu} \xi^{\nu} T_{\mu\nu}\,,
\end{equation}
from which we easily find
\begin{equation}\label{eq:energy_hol}
    \delta E = 2\pi \int_{A} d^{d-1} x \left(\frac{R^{2}-\vec{x}^{2}}{2R}\right) \delta \langle T_{00} \rangle\,.
\end{equation}
On the other hand, $\delta \langle T_{\mu\nu} \rangle$ can be related to the metric perturbation $H_{\mu\nu}$ by holographic renormalization:
\begin{equation}
    T_{\mu\nu}{(x,t)} = \frac{d}{16\pi G} H_{\mu\nu} {(z=0,x,t)}\,.
\end{equation}
Plugging back into \eqref{eq:energy_hol}, we readily see that, indeed, $\delta E\equiv \int_{\partial \Sigma_\infty}\chi $.

Next, we move on to discuss the Gauss-Bonnet case. Intriguingly enough, the linearized equation of motion in AdS is exactly {\it the same} as in Einstein gravity \cite{Boulware:1985wk}. In particular, we still want the perturbation $H_{\mu\nu}$ to be traceless and divergence-free. To keep the algebra manageable, we work in $d=4$  and consider a particular perturbation of the form:
\begin{equation}\label{eq:GB_metric_pert}
ds^{2} = \frac{L^{2}}{z^{2}} \bigg(dz^{2} + (-1 + H z^{4})dt^{2} + \left(1+ \frac{Hz^{4}}{3} \right)(dx_{1}^{2}+dx_{2}^{2}+dx_{3}^{2}) \bigg),
\end{equation}
where $H$ is a constant. Next, we compute the form $\chi$, but first we need $\delta \textbf{Q}$ and $\bf{\Theta}$ under the above perturbation. We find for the variation of the Noether charge:
\begin{equation}
    \delta \textbf{Q}|_{\Sigma} = \sum_{i < j} \delta \textbf{Q}_{zij} dz \wedge dx^{i} \wedge dx^{j} + \delta \textbf{Q}_{123} dx^{1} \wedge dx^{2} \wedge dx^{3}\,,
\end{equation}
with
\begin{equation}\label{dQzijGB}
    \delta \textbf{Q}_{zij} = \frac{HLz}{12GR}\epsilon_{ijk}x^{k}(L^{2}-448G\pi\alpha)
\end{equation}
and
\begin{equation}\label{dQ123GB}
\delta \textbf{Q}_{123} = -\frac{HL}{8GR}\left[L^{2}(2R^{2}-2\vec{x}^{2}-z^{2}) - 64\pi G\alpha (2R^{2} - 2\vec{x}^{2} - 7z^{2}) \right]
\end{equation}
while the boundary term vanishes (see Appendix \ref{app:theta0}):
\begin{equation}\label{eqn427}
{\bf{\Theta}} = 0\,.
\end{equation}
Therefore, the form $\chi$ coincides with the variation of $\textbf{Q}$. As usual, the integral of $\chi$ over the bifurcation surface yields $\delta S$. In this example, it might be worthwhile to see this explicitly. The restriction of $\chi$ to the bifurcation surface is:
\begin{equation}
\chi \bigg|_{\partial \Sigma_{h}} = -\frac{HL}{24GR} \left[L^{2}(3R^{2}-\vec{x}^{2}) + 64G\pi\alpha (15R^{2}-29\vec{x}^{2}) \right] dx^{1} \wedge dx^{2} \wedge dx^{3}\,.
\end{equation}
Integrating $\chi$ over the bifurcation surface then gives:
\begin{equation}\label{chionbifurcationGB}
\int_{\partial \Sigma_{h}} \chi = -\frac{2 \pi  H L R^4 \left(L^2-64 \pi  \alpha  G\right)}{15 G}\,.
\end{equation}
On the other hand, let us compute the change in the area of the Ryu-Takayanagi surface due to $H$. From the modified area functional (\ref{GBRTaction}), the shift in the entanglement entropy is given by:
\begin{equation}\label{deltaSGB}
    \delta S = \frac{1}{4G} \int_{M} d^{3}x [\delta(\sqrt{h}) + 32\pi G \alpha \delta (\sqrt{h} \mathcal{R})]\,.
\end{equation}
with
\begin{equation}
\delta{(\sqrt{h})} = H\frac{L^{3}}{6R}(3R^{2}-\vec{x}^{2})\,,
\end{equation}
and
\begin{equation}
\delta{(\sqrt{h} \mathcal{R})} = \frac{HL}{3R} (15R^{2} - 29\vec{x}^{2})\,.
\end{equation}
If we plug the two equations above into equation (\ref{deltaSGB}) and integrate to obtain the variation of entanglement entropy, we then find
\begin{equation}
\delta S_{EE} = \frac{2 \pi  H L R^4 \left(L^2-64 \pi \alpha G\right)}{15 G}\,.
\end{equation}
Comparing with the integral of $\chi$ over the bifurcation surface given in (\ref{chionbifurcationGB}), we find agreement:
\begin{equation}\label{eq:metric_pert_GB_hor_term}
\int_{\partial \Sigma_{h}} \chi = -\delta S_{EE}\,.
\end{equation}
Finally, the restriction of $\chi$ to the boundary is:
\begin{equation}
\chi|_{\partial \Sigma_{\infty}} = -\frac{HL}{4GR}(L^{2}-64G\pi\alpha) (R^{2}-\vec{x}^{2}) dx^{1} \wedge dx^{2} \wedge dx^{3}\,.
\end{equation}
 Integrating this over the boundary yields:
\begin{equation}\label{deltaE}
\delta E = \frac{2 \pi  H L R^4 \left(L^2-64 \pi \alpha G\right)}{15 G}\,.
\end{equation}
Comparing with the integral over the horizon, we find agreement. Of course, the result for $\delta E$ obtained here is consistent with the holographic stress-energy tensor computed from holographic renormalization in Gauss-Bonnet theory.

\section{Proof of eqs. (\ref{eqn325}) and (\ref{eqn427}) }\label{app:theta0}
First, we show that the symplectic potential current vanishes under a perturbation of $L$ in Einstein gravity. A variation of $L$  changes the metric in the following way.
\begin{equation}
    \delta g_{tt} = -\delta g_{zz} = -\frac{2L\delta L}{z^{2}}\,,
\end{equation}
\begin{equation}
    \delta g_{ij} = \frac{2L \delta L}{z^{2}} \delta_{ij}\,.
\end{equation}
The nonzero Christoffel symbols of the $(d+1)$-dimensional Poincar\'{e} patch are:
\begin{equation}
    \Gamma^{t}_{tz} = \Gamma^{z}_{tt} = \Gamma^{z}_{zz} = -\frac{1}{z}\,,
\end{equation}
\begin{equation}
    \Gamma^{i}_{zj} = -\delta^{i}_{j} \frac{1}{z}\,,
\end{equation}
\begin{equation}
    \Gamma^{z}_{ij} = \delta_{ij} \frac{1}{z}\,.
\end{equation}
In order to show that the symplectic potential current vanishes, we will show that the following two quantities vanish:
\begin{equation}
    \Theta_{(1)}^{a} = g^{ac}g^{bd} \nabla_{b} \delta g_{cd}\,,
\end{equation}
\begin{equation}
    \Theta_{(2)}^{a} = g^{ac}g^{bd} \nabla_{c} \delta g_{bd}\,.
\end{equation}
Consider first the second quantity. We can recast it as:
\begin{equation}
\Theta_{(2)}^{a} = g^{ac} \partial_{c} (g^{bd} \delta g_{bd})\,.
\end{equation}
But the quantity in parentheses is can be found to be:
\begin{equation}
g^{bd} \delta g_{bd} = \frac{2\delta L}{L}(d+1)\,.
\end{equation}
In particular this quantity has no spacetime dependence, and therefore any partial derivative of this quantity vanishes, and we find:
\begin{equation}
\Theta_{(2)}^{a} = 0\,.
\end{equation}
Next, consider the quantity ${\bf{\Theta}}_{(1)}^{a}$. A lengthy but straightforward calculation using the Christoffel symbols listed above reveals that this quantity also vanishes for each choice of $a$ ($a = t, z, x^{i}$). Thus we find that the symplectic potential current vanishes under variations of $L$.
\begin{equation}
 {\bf{\Theta}}=0\,.
 \end{equation}
This proves equation (\ref{eqn325}). Next, we show that the symplectic potential current also vanishes under a metric perturbation $H_{\mu\nu}$ in the Einstein-Gauss-Bonnet theory. Recall that the general expression for Einstein-Gauss-Bonnet, given in (\ref{ThetaGB}), is not proportional to the sum of $\Theta_{(1)}$ and $\Theta_{(2)}$. However, when evaluated on the AdS background, the result is proportional to this sum (see equation \ref{ThetaGBAdS}). Thus we will show again that both $\Theta_{(1)}$ and $\Theta_{(2)}$ vanish. The metric changes in the following way due to $H_{\mu\nu}$:
\begin{equation}
\delta g_{\mu\nu} = L^{2}z^{d-2}H_{\mu\nu}{(x^{\lambda})}\,.
\end{equation}
In this case, we find:
\begin{equation}
\Theta^{a}_{(2)} = g^{ac}\partial_{c} \left(z^{d} \eta^{\mu\nu} H_{\mu\nu} \right)\,,
\end{equation}
\begin{equation}
\Theta^{t}_{(1)} = \frac{z^{d+2}}{L^{2}} (\partial_{t}H_{tt} - \partial^{i}H_{ti})\,,
\end{equation}
\begin{equation}
\Theta^{z}_{(1)} = 0\,,
\end{equation}
\begin{equation}
\Theta^{i}_{(1)} = \frac{z^{d+2}}{L^{2}} \partial^{k} H^{i}_{k}\,.
\end{equation}
In particular, for the perturbation in Section 4.2, all three equations above vanish. This proves equation (\ref{eqn427}).


%
\end{document}